\documentclass[amsmath,amssymb,aps,showkeys,preprint, %
floatfix,groupedaddress,superscriptaddress,amsmath,amssymb, 
nofootinbib,
prd]{revtex4-1}

\usepackage{dcolumn}
\usepackage{amsmath,amssymb}
\usepackage[breaklinks = true,colorlinks = true,%
linkcolor = blue,citecolor = blue]{hyperref}
\usepackage{graphicx}
\usepackage{xcolor}
\usepackage{ascmac}
\usepackage{slashed}

\makeatletter
\let\@makefntextOrig\@makefntext
\def\@makefntext#1{\@makefntextOrig{\baselineskip=7.2pt #1}}
\makeatother

\def\ee{\end{eqnarray}}

\def\p{\partial}

\def\=:{=\hspace{-.7em}\raisebox{1.1ex}{.}\hspace{.1em}\raisebox{-0.2ex}{.} }

\def\ee{\end{eqnarray}}

\def\p{\partial}

\def\=:{=\hspace{-.7em}\raisebox{1.1ex}{.}\hspace{.1em}\raisebox{-0.2ex}{.} }

\newcommand {\beq}{\begin{eqnarray}}
\newcommand {\eeq}{\end{eqnarray}}
\newcommand {\non}{\nonumber\\}

\newcommand {\1}[1]{\frac{1}{#1}}

\newcommand {\ph}{\varphi}

\newcommand {\del}{\partial}

\newcommand {\tr}{{\rm tr}\,}

\newcommand {\kahler}{K\"ahler }

\begin{document}


\title{
Relations among topological solitons
}


\author{Muneto Nitta}

\affiliation{
Department of Physics, and Research and Education Center for Natural 
Sciences, Keio University, Hiyoshi 4-1-1, Yokohama, Kanagawa 223-8521, Japan\\
}


\date{\today}
\begin{abstract}
We clarify relations among topological solitons in various dimensions:
  a domain wall, non-Abelian vortex, magnetic monopole 
  and Yang-Mills instanton, 
  together with a (non-Abelian) sine-Gordon soliton, baby Skyrmion (lump), and Skyrmion.
  We construct a composite configuration 
consisting of a domain wall, vortex, magnetic monopole and Yang-Mills instanton (wall-vortex-monopole-instanton) 
using the effective theory technique or moduli approximation. 
Removing some solitons from such a composite, 
we obtain all possible composite solitons in the form of 
solitons within a soliton, including all the previously known configurations,
yielding relations among topological solitons. 

\end{abstract}

\maketitle

\section{Introduction}
Topological solitons and instantons are ubiquitous in nature: 
they appear and play significant roles in 
quantum field theories 
\cite{Rajaraman:1987,Manton:2004tk, Shnir:2005vvi,Vachaspati:2006zz,Dunajski:2010zz,Weinberg:2012pjx,Shnir:2018yzp},
supersymmetric field theories 
\cite{Tong:2005un,*Tong:2008qd,Eto:2006pg,Shifman:2007ce,*Shifman:2009zz},
cosmology 
\cite{Kibble:1976sj,*Kibble:1980mv,Vilenkin:1984ib,*Hindmarsh:1994re,*Vachaspati:2015cma,Vilenkin:2000jqa}, 
QCD 
\cite{Eto:2013hoa},
and various condensed matter systems 
\cite{Mermin:1979zz}
such as  
helium-3 superfluids \cite{Volovik:2003fe}, 
superfluids 
\cite{Svistunov:2015},
Josephson junctions in superconductors \cite{Ustinov2015}, 
nonlinear media 
\cite{Pismen,*Bunkov:2000}, and 
Bose-Einstein condensates (BECs) of ultracold atomic gasses 
\cite{Kawaguchi:2012ii}. 
Formations and dynamics of topological solitons have been 
discussed in various contexts.
Among them, 
 topological defects are ubiquitously formed 
 during phase transitions 
(the Kibble-Zurek mechanism) \cite{Kibble:1976sj,*Kibble:1980mv,Zurek:1985qw,*Zurek:1996sj}, 
and thus have been studied extensively in cosmology and 
condensed matter physics. 
Dynamics of topological solitons are mostly studied numerically. 
When exact or approximate soliton solutions are available, 
their low-energy dynamics can be analytically described by 
the moduli approximation \cite{Manton:1981mp}
(for solitons with supersymmetry, see Ref.~\cite{Eto:2006uw}).

\begin{table}[h]
\begin{tabular}{|c|cccc|cc|}
\hline
Topological solitons & Type & {\tiny Codim} & SSB & Homotopy & Example & Single soliton moduli 
\\ \hline\hline
Yang-Mills instantons & Gauge  & 4 &  Non & $\pi_3(G)$ & 
$\pi_3 [SU(N)] \simeq {\mathbb Z}$ 
& ${\mathbb R}^4 \times {\mathbb R}^+ \times 
\frac{SU(N)}{SU(N-2)\times U(1)}$ 
\\ \hline
Magnetic monopoles   & Defect & 3 &  Gauge & $\pi_2(G/H)$
&  $\pi_2 \left[\frac{SU(N)}{U(1)^{N-1}}\right] \simeq {\mathbb Z}^{N-1}$ 
& ${\mathbb R}^3 \times U(1)$
\\
Vortices                    & Defect & 2 &  Gauge & $\pi_1(G/H)$ & 
$\pi_1 [U(N)] \simeq {\mathbb Z}$ 
& ${\mathbb R}^2 \times {\mathbb C}P^{N-1}$ 
\\
Kinks, Domain walls    & Defect & 1 &  Global & $\pi_0(G/H)$ & 
$\pi_0({\mathbb Z}_N) \simeq {\mathbb Z}_N$ & ${\mathbb R} \times U(N)$ 
\\
\hline
Skyrmions                 & Texture& 3 & Global & 
$\pi_3(G/H)$ &  $\pi_3 [SU(N)]\simeq {\mathbb Z}$  
& ${\mathbb R}^3 \times \frac{SU(N)}{SU(N-2)\times U(1)}$ 
\\
Baby Skyrmions(Lumps)    & Texture& 2 & Global & $\pi_2(G/H)$ & 
$\pi_2[{\mathbb C}P^{N-1} ]\simeq {\mathbb Z}$  
& ${\mathbb R}^2 \times U(1)$ 
 $({\mathbb R}^2\times {\mathbb C}^*)$ \\
Sine-Gordon solitons  & Texture& 1 & Global & $\pi_1(G/H)$ & 
 $\pi_1 [U(N)] \simeq {\mathbb Z}$
 & ${\mathbb R}\times {\mathbb C}P^{N-1}$  \\
 Hopfions & Texture& 3 & Global &  $\pi_3(S^2)$ & $\pi_3(S^2)\simeq {\mathbb Z}$ & ${\mathbb R}^3 \times U(1)$\\
 \hline
 \end{tabular} 
  \caption{\label{table:topological-solitons}
  ``Type'' implies the types of topological solitons classified  
  into topological gauge structures, defects, or textures, 
  and ``Codim'' implies codimensions on which soliton configurations depend.
  Topological gauge structures imply topological solitons in pure gauge theories in an $n$-dimensional space ${\mathbb R}^n$, 
  classified by a homotopy group $\pi_{n-1} (G)$ 
of a map from the boundary $S^{n-1}$ of ${\mathbb R}^n$ 
to a gauge group $G$. 
  Topological defects arise in SSB $G\to H$ of either gauge or global symmetry $G$, 
  classified by a homotopy group $\pi_{n-1} (G/H)$ 
  of a map from the boundary $S^{n-1}$ to the OPM $G/H$. 
  Topological textures are classified by a homotopy group 
  $\pi_n (G/H)$ of a map from the whole $n$-dimensional space 
  ${\mathbb R}^n$ to 
  the OPM $G/H$ 
  with one point compactification of the base space ${\mathbb R}^n$ to $S^n$
  (justified by the assumption that 
   the boundary is mapped to the same point).
   ``Example'' gives typical examples, 
   and ``Single soliton moduli'' denotes moduli space 
   of a single soliton in such examples.
  }
\end{table}

Topological solitons can be classified into 
topological defects, topological textures, 
and topological gauge structures(instantons).
Topological defects are formed 
 due to the aforementioned Kibble-Zurek mechanism 
when  
spontaneous symmetry breaking (SSB) 
$G \to H$
of gauge or global symmetry $G$ 
 occurs.
 Nontrivial homotopy groups $\pi_{n-1}(G/H)$  
 of the order parameter 
manifold (OPM) $G/H$ admit the existence of 
topological defects of codimension $n$ 
(particle-like in an $n$ dimensional space ${\mathbb R}^d$), 
which can be surrounded by $S^{n-1}$ 
at the boundary of ${\mathbb R}^n$.  
One of the remarkable features of topological defects is the fact that 
the unbroken symmetry $H$ is partially or fully recovered 
around their cores. 
On the other hand, 
  topological textures are classified by a homotopy group 
  $\pi_n (G/H)$ as a map from the whole $n$-dimensional space 
  ${\mathbb R}^n$ to 
  the OPM $G/H$ 
  with one point compactification of the base space ${\mathbb R}^n$ to $S^n$,
  which can be justified by the assumption that 
   the boundary is mapped to the same point.
Gauge structures imply nontrivial pure gauge field configurations 
without SSB classified by 
a homotopy group $\pi_{n-1}(G)$  of the gauge group $G$ on
an $n$-dimensional space ${\mathbb R}^n$.  
Taking into account their dimensionality up to four space-time dimensions, 
there are following eight topological solitons,  
as summarized in Table \ref{table:topological-solitons}:

Yang-Mills instantons are solutions to 
self-dual Yang-Mills equations \cite{Belavin:1975fg} 
 which are of codimension four (point-like in the Euclidean 
 space ${\mathbb R}^4$)
 and can be particle-like in $d=5+1$ dimensions. 
The (framed) moduli space of a single $SU(N)$ instanton is
${\mathbb R}^4 \times {\mathbb R}^+ \times 
\frac{SU(N)}{SU(N-2)\times U(1)}$. 
For multi-instanton configurations and their moduli space,
the Atiyah-Drinfeld-Hitchin-Manin (ADHM) construction is available 
\cite{Atiyah:1978ri,*Corrigan:1983sv}. 
Instantons have important applications 
to determine  
non-perturbative effects in quantum field 
theories in four space-time dimensions, 
in particular, with supersymmetry 
\cite{Seiberg:1994rs, *Seiberg:1994aj,*Dorey:2002ik,*Nekrasov:2002qd}.

`t Hooft-Polyakov magnetic monopoles are 
particle-like topological defects in $d=3+1$ 
\cite{tHooft:1974kcl,*Polyakov:1974ek}
(see Refs.~\cite{Shnir:2005vvi,Preskill:1984gd,*Weinberg:2006rq,*Konishi:2007dn} as a review)  
and instanton-like in $d=2+1$. 
The simplest SSB $SU(2) \to U(1)$
admits a  nontrivial second homotopy group  
$\pi_2 [SU(2)/U(1)] \simeq {\mathbb Z}$,  
supporting monopoles. 
This  
can be generalized to an $SU(N)$ gauge group 
maximally broken by the adjoint Higgs field as  
$SU(N) \to U(1)^{N-1}$,  
with $\pi_2 [SU(N)/U(1)^{N-1}] \simeq {\mathbb Z}^{N-1}$.
In the
Bogomol'nyi-Prasad-Sommerfield (BPS) 
limit, analytic solutions are available \cite{Bogomolny:1975de,*Prasad:1975kr}, 
for which 
the Nahm construction offers solutions of multiple BPS monopoles 
and their moduli space \cite{Nahm:1979yw}.
The moduli spaces of two monopoles \cite{Atiyah:1985dv} 
and well-separated multiple monopoles \cite{Gibbons:1995yw} 
are available.
There are several alternative methods such as 
the Donaldson's rational map 
\cite{Donaldson:1984ugy}.
When the $SU(N)$ gauge symmetry is not maximally broken 
by the adjoint Higgs field, 
there remains a partially unbroken 
non-Abelian gauge symmetry in the vacuum.
In this case, monopoles are non-Abelian monopoles 
\cite{
Goddard:1976qe,*Weinberg:1979zt,*Weinberg:1982ev,
*Auzzi:2004if}. 
In high-energy phenomenology and cosmology, 
grand unified theories (GUTs) always admit GUT monopoles 
(the monopole problem) because of 
$\pi_2[G/(SU(3)\times SU(2) \times U(1))] \simeq {\mathbb Z}$
\cite{Dokos:1979vu,*Lazarides:1980va,*Preskill:1979zi}.

Vortices or cosmic strings are topological defects of codimension two,
which are string-like in $d=3+1$, particle-like in $d=2+1$, 
and instanton-like in $d=1+1$ 
(see Refs.~\cite{Vilenkin:2000jqa,Manton:2004tk} as a review).
The simplest case is 
Abrikosov-Nielsen-Olesen (ANO) vortices 
(magnetic flux tubes)
\cite{Abrikosov:1956sx,*Nielsen:1973cs} 
supported by $\pi_1[U(1)] \simeq {\mathbb Z}$  
when a $U(1)$ gauge symmetry is spontaneously broken 
relevant for superconductors. 
Global analogues are 
global vortices
having logarithmically divergent tension 
in infinite space, 
examples of which are given by axion strings 
\cite{Vilenkin:1982ks,*Kawasaki:2013ae}
in cosmology, 
and superfluid vortices 
in superfluids or atomic BECs 
in condensed matter physics.
In an SSB of both local and global symmetries, 
semilocal strings are known 
\cite{Vachaspati:1991dz, *Achucarro:1999it}. 
Some time ago, 
non-Abelian vortices containing non-Abelian magnetic fluxes inside them were discovered \cite{Hanany:2003hp,*Auzzi:2003fs},   
and since then they have been extensively studied in various contexts  
(see Refs.~\cite{Tong:2005un,*Tong:2008qd,Eto:2006pg,Shifman:2007ce,*Shifman:2009zz} as a review).
The moduli space of a single $U(N)$ vortex 
is ${\mathbb C} \times {\mathbb C}P^{N-1}$, 
and thus the low-energy effective world-sheet theory 
is a two-dimensional ${\mathbb C}P^{N-1}$ model, 
explaining similarities between 
four dimensional gauge theories and two-dimensional sigma models. 
Multiple vortex solutions and their moduli space are available
by the moduli matrix method  
\cite{Eto:2005yh,*Eto:2006cx,*Eto:2006db} 
and half-ADHM construction \cite{Eto:2006pg}.
The moduli space metric can be calculated for 
well-separated vortices 
\cite{Fujimori:2010fk} 
yielding the low-energy dynamics \cite{Eto:2011pj}, 
while coincident vortices can be described group theoretically 
\cite{Eto:2010aj}.
These vortices can be generalized to 
non-Abelian semilocal strings 
\cite{Shifman:2006kd,*Eto:2007yv} 
and those of arbitrary gauge groups 
\cite{Eto:2008yi,*Eto:2009bg}.
In high energy phenomenology, 
similar non-Abelian vortices are present 
in high density QCD (color superconductors) \cite{Balachandran:2005ev,*Nakano:2007dr,*Nakano:2008dc,*Eto:2009kg,*Eto:2009bh,*Eto:2009tr,*Alford:2016dco} 
(see Ref.~\cite{Eto:2013hoa} as a review, 
and Refs.~\cite{Alford:2018mqj,*Chatterjee:2018nxe} for recent developments)  
relevant for neutron star cores 
and in two-Higgs doublet model (2HDM) \cite{Dvali:1993sg,*Eto:2018hhg,*Eto:2018tnk,*Eto:2021dca}.
On the other hand, electroweak $Z$-strings in the Standard Model also have non-Abelian magnetic fluxes 
\cite{Nambu:1977ag,*Vachaspati:1992fi} 
but they are nontopological and unstable.

Finally, the simplest topological defects are domain walls or kinks 
of codimension one 
(see Refs.~\cite{Manton:2004tk,Vachaspati:2006zz,Shnir:2018yzp} as a review).
While the $\phi^4$ kink is the simplest example, 
one of the more interesting examples is
the ${\mathbb C}P^1$ domain wall 
 (or those in a $U(1)$ gauge theory with two complex Higgs scalar fields) 
\cite{Abraham:1992vb,*Abraham:1992qv,*Arai:2002xa,*Arai:2003es}.
These are relevant for ferromagnets 
and supersymmetric theories.
These can be generalized to
${\mathbb C}P^{N-1}$ domain walls (or those in a $U(1)$ gauge theory with $N$ complex Higgs scalar fields) 
\cite{Gauntlett:2000ib,*Tong:2002hi}. 
The moduli space of a single kink is ${\mathbb R} \times U(1)$ 
if masses are nondegenerate
(for degenerate masses \cite{Eto:2008dm}).
Further generalizations can be found as
Grassmann domain walls 
(or a $U(N)$ gauge theory) \cite{Isozumi:2004jc,*Isozumi:2004va,
*Isozumi:2004vg,*Eto:2004vy,Hanany:2005bq}, 
those with 
generic $U(1)$ charges \cite{Eto:2005wf},
and those in nonlinear sigma models 
with the other target spaces 
\cite{Arai:2009jd,*Arai:2011gg,*Lee:2017kaj,*Arai:2018tkf}.
Among all the cases, a particularly important case in this paper is given by 
a non-Abelian domain wall
\cite{Shifman:2003uh,Eto:2005cc,Eto:2008dm}
that 
carries non-Abelian moduli ${\mathbb R} \times U(N)$, 
and thus its low-energy effective theory is a $U(N)$ 
chiral Lagrangian (principal chiral model), 
or the Skyrme model if a four derivative term is taken into account 
\cite{Eto:2005cc}. 

Apart from instantons and topological defects, 
another class of topological solitons is given by topological textures such as Skyrmions. 
Skyrmions \cite{Skyrme:1961vq,*Skyrme:1962vh} 
are topological textures of the codimension three characterized by 
$\pi_3[SU(2)]\simeq {\mathbb Z}$ in the chiral Lagrangian of pions 
with a four derivative (Skyrme) term, 
which were proposed to describe baryons 
\cite{Witten:1983tx} 
(see Refs.~\cite{Manton:2004tk,Shnir:2018yzp} as a review). 
The moduli space of a single $SU(N)$ Skyrmion is 
${\mathbb R}^3 \times 
\frac{SU(N)}{SU(N-2)\times U(1)}$. 
While analytic solutions are not available, 
there are some proposals to give approximate configurations, 
such as 
the Atiyah-Manton construction 
based on Yang-Mills instantons
 \cite{Atiyah:1989dq,*Atiyah:1992if} 
 and  
rational map ansatz \cite{Houghton:1997kg,Ioannidou:1999mf}.

Baby Skyrmions or planar Skyrmions  
are $2+1$ dimensional analogues of Skyrmions 
characterized by the second homotopy group $\pi_2(S^2)\simeq {\mathbb Z}$, 
typically present in a $O(3)$ sigma model with a potential term 
and a four derivative term \cite{Piette:1994mh,*Piette:1994ug} 
(see Refs.~\cite{Manton:2004tk,Shnir:2018yzp} as a review).
They are string-like in $d=3+1$, particle-like in $d=2+1$ 
and instanton-like in $d=1+1$. 
The case without a four-derivative term was known earlier 
as lumps  or sigma model instantons \cite{Polyakov:1975yp}.
The moduli space of a single lump (baby Skyrmion)
 is ${\mathbb C} \times {\mathbb C}^*$ (${\mathbb C} \times U(1)$).
 In condensed matter physics, 
these solitons are simply called Skyrmions
and are relevant in various systems such as 
quantum Hall effects \cite{Ezawa} and ferromagnets.  
In particular there are great interests in chiral magnets 
because of experimental finding of a Skyrmion lattice in chiral magnets \cite{Muhlbauer:2009,*Yu:2010,*Han:2010by,*Lin:2014ada}
(see Refs.~\cite{Barton-Singer:2018dlh,*Schroers:2019gbe,*Ross:2020hsw} for analytic studies of Skyrmions and their lattice in chiral magnets). 
The same Skyrmions can be present also in chiral liquid crystals \cite{Foster:2019rbd}.
The moduli space of a single Skyrmion is only ${\mathbb C}$.

Finally, the lowest dimensional analogues of Skyrmions are 
sine-Gordon solitons 
\cite{Perring:1962vs}   
(and 
double sine-Gordon solitons \cite{PhysRevB.27.474,*Gani:2017yla}), 
characterized by the first homotopy group 
$\pi_1[U(1)] \simeq {\mathbb Z}$ 
(see Refs.~\cite{Manton:2004tk,Shnir:2018yzp} as a review).
These solitons are of codimension one: 
planar in $d=3+1$, string-like in $d=2+1$, and particle-like in $d=1+1$.
They are 
similar to domain walls 
but are distinct from them in the sense that   
the vacua far from solitons are identical 
as the other Skyrmions, 
in contrast to domain walls separating distinct vacua.
In cosmology, a sine-Gordon soliton is 
attached to an axion string 
(if the domain wall number is one) 
\cite{Vilenkin:1982ks,*Kawasaki:2013ae}. 
Double sine-Gordon solitons also appear 
in various context such as 
2HDM in high energy physics~\cite{Eto:2018hhg,*Eto:2018tnk}. 
In condensed matter physics, 
(double) sine-Gordon solitons are present in a Josephson junction 
(an insulator sandwiched by two superconductors)  \cite{USTINOV1998315,Ustinov2015}. 
A lattice of (double) sine-Gordon solitons is the ground state 
in chiral magnets (in a certain parameter region), which is called a chiral soliton lattice  \cite{togawa2012chiral,*togawa2016symmetry,
*PhysRevB.97.184303,*PhysRevB.65.064433,*Ross:2020orc}
(see Ref.~\cite{KISHINE20151} as a review).
Recently, in high energy physics, 
such chiral soliton lattices are found to be realized  
as the ground states 
in QCD under rapid rotation \cite{Huang:2017pqe,*Nishimura:2020odq} or
in the strong magnetic field \cite{Brauner:2016pko}. 
Among various extensions of sine-Gordon solitons, 
particularly important ones in this study are 
non-Abelian sine-Gordon solitons supported by 
$\pi_1[U(N)] \simeq {\mathbb Z}$ 
\cite{Nitta:2014rxa,Eto:2015uqa}.
In QCD, 
a single non-Abelian sine-Gordon soliton 
can be bounded by global non-Abelian strings 
\cite{Balachandran:2002je,*Nitta:2007dp,*Nakano:2007dq,
*Eto:2009wu,*Eto:2013bxa,Eto:2013hoa}
and is stretched between two chiral non-Abelian strings  
\cite{Eto:2021nle}. 
The moduli space of a single $U(N)$ sine-Gordon soliton  
is ${\mathbb R} \times {\mathbb C}P^{N-1}$,
and thus the low energy theory is 
the ${\mathbb C}P^{N-1}$ model, as the case of non-Abelian vortices.
Such solitons can exist in a non-Abelian extension of 
a Josephson junction (two color superconductors separated by an insulator or a domain wall) \cite{Nitta:2015mma,Nitta:2015mxa}.
More interestingly, a chiral soliton lattice of 
non-Abelian sine-Gordon solitons 
is found to be the ground state of QCD 
under rapid rotation 
in a wide range of the parameter region \cite{Eto:2021gyy} 
(instead of Abelian chiral soliton lattices  \cite{Huang:2017pqe,*Nishimura:2020odq}).

The last topological textures are Hopfions 
characterized by a Hopf map $\pi_3(S^2) \simeq {\mathbb Z}$ 
\cite{Faddeev:1996zj} 
(see Refs.~\cite{Radu:2008pp,Shnir:2018yzp} as a review).
Hopfions are present in an $O(3)$ sigma model 
with a four derivative term, called the Faddeev-Skyrme model. 
Typically these solitons are string-like and are linked 
\cite{Battye:1998pe,*Battye:1998zn,*Kobayashi:2013xoa}.
In condensed matter physics,  
Hopfions were suggested in $^3$He superfluids  
\cite{Volovik:1977} and
 superconductors \cite{Babaev:2001zy,*Rybakov:2018ktd}, 
and  have been experimentally confirmed 
 in a spinor BEC 
(but are unstable without a four derivative term) 
\cite{Kawaguchi:2008xi,*Kawaguchi:2010mu,*Hall:2016,*Ollikainen:2019dyh}, 
liquid crystals \cite{PhysRevLett.110.237801,
*Ackerman:2015,*Ackerman:2017,*Ackerman:2017b,
*Tai:2018,*Tai:2019}, 
and magnetic materials \cite{Kent:2021}.

\begin{figure}[h]
\begin{center}
\includegraphics[width=0.5\linewidth,keepaspectratio]{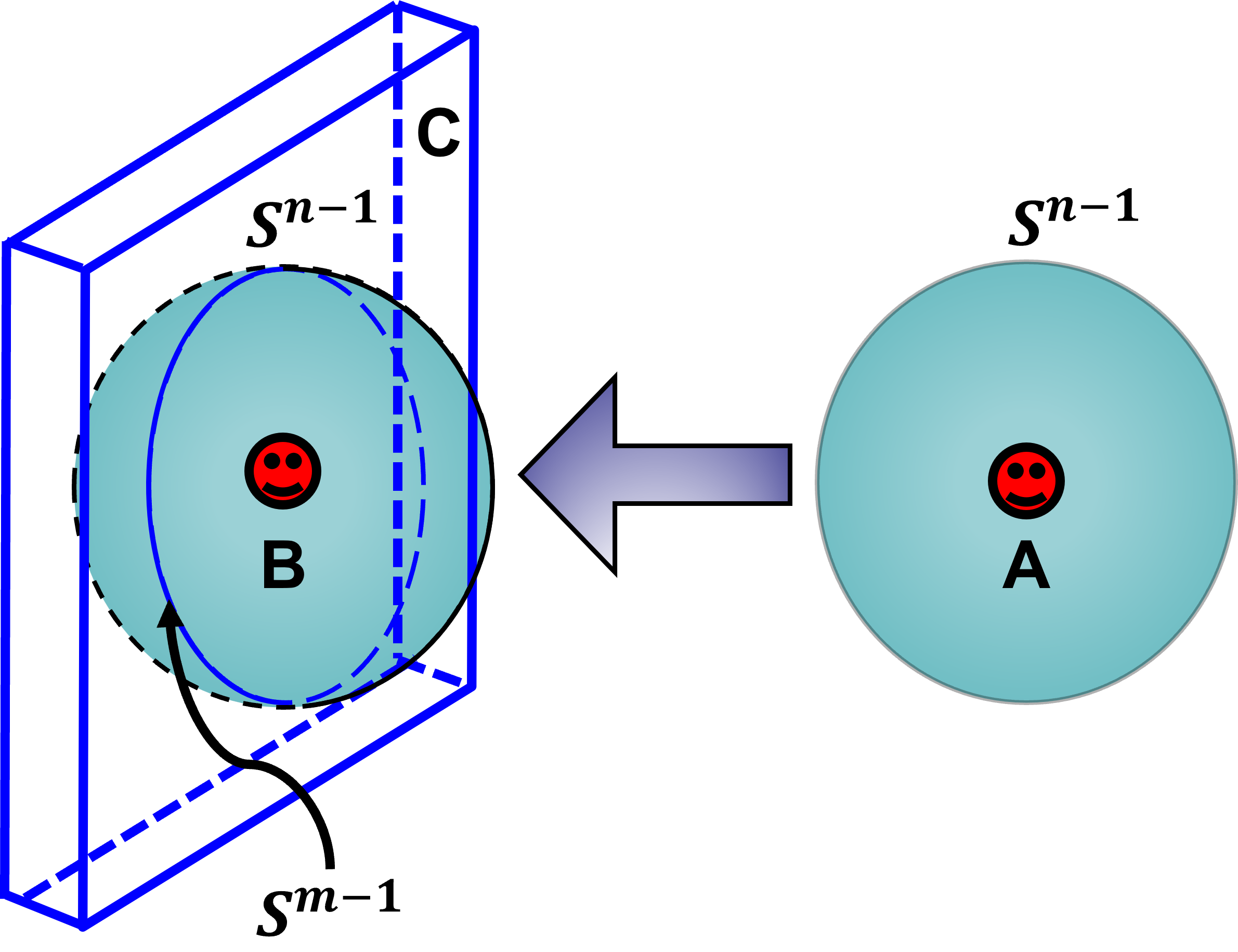}
\caption{A relation of topological solitons.
When a daughter ``A'' of codimensions $n$, surrounded by $S^{n-1}$ in the bulk, is absorbed 
into a mother ``C'' it becomes ``B''.  
The effective theory of the moduli fields of C admits a topological soliton B surrounded by $S^{m-1}$, which has the same topological charge with A in the bulk.
Thus, A and B can be identified. 
\label{relation}
}
\end{center}
\end{figure}

As we have seen,
various topological solitons appear in diverse subjects. 
It is, however, usually the case that 
each soliton is studied individually, 
and a unified understanding is yet to be clarified.  
The purpose of this paper is to present 
connections among all kinds of topological solitons 
with a help of composite solitons 
in the form of ``solitons within a soliton,'' 
where a
{\it daughter} soliton is trapped inside a {\it mother} soliton,
as schematically shown in Fig.~\ref{relation}.\footnote{
There are also other types of composite solitons: 
solitons ending on a soliton or soliton junctions, 
such as a non-Abelian vortex ending on a monopole 
\cite{Auzzi:2003em,*Eto:2006dx}, 
electroweak $Z$-string ending on a Nambu monopole 
\cite{Nambu:1977ag}, 
axion domain wall(s) ending on an axion string \cite{Vilenkin:1982ks,*Kawasaki:2013ae},  
and
axial (chiral) domain walls(s) ending on an axial vortex 
\cite{Balachandran:2002je,*Nitta:2007dp,*Nakano:2007dq,
*Eto:2009wu,*Eto:2013bxa,Eto:2013hoa,Eto:2013bxa,Eto:2021nle}.
In addition, vortices ending on a domain wall or stretched between 
domain walls, and domain wall junctions, 
which can be BPS, are known. 
We do not consider these in this paper. 
}
Thus, the mother soliton should have larger world-volume dimensions 
(less codimensions) than the daughter soliton  
(see Appendix \ref{sec:same-dim} for the exceptional cases in which  
mother and daughter solitons have the same dimensions 
where a single daughter soliton must be split into fractional 
solitons).
In general, 
the mother soliton possesses {\it moduli parameters} 
or collective coordinates, 
that is, a set of solutions with the same energy 
contains free parameters.
Then, the  low-energy effective world-volume theories 
of the mother solitons can be constructed by 
the moduli approximation \cite{Manton:1981mp}
(with supersymmetry for BPS solitons \cite{Eto:2006uw}), 
where the moduli are promoted to fields on 
the world-volume.
It is usually a nonlinear sigma model  
whose target space is the moduli space.

Let the codimension of the daughter soliton 
be $n$ in the bulk 
so that the daughter soliton is surrounded by $S^{n-1}$ 
as A in Fig.~\ref{relation}.
When it is absorbed into the mother soliton C,   
it can be expressed by a soliton (at B) 
in the effective world-volume theory 
of the mother soliton \footnote{
On the other hand, 
a defect in the effective theory of 
the mother soliton represents  
a defect ending on the mother defect. 
Examples are given by 
D-brane solitons 
\cite{Gauntlett:2000de,*Shifman:2002jm,*Isozumi:2004vg,
Auzzi:2005yw,Eto:2008mf}
(see also \cite{Bowick:2003au}).
Higher dimensional generalizations can be found in 
Ref.~\cite{Gudnason:2014uha}. 
We do not consider these cases in this paper. 
}.   
In this case, 
the daughter soliton inside the mother soliton 
is surrounded by $S^{m-1}$, 
which is the intersection of 
$S^{n-1}$ at B and the world-volume of the mother soliton,
where 
$m$ can be expressed by 
$n$ and
the codimension $c$ of the mother soliton 
as $m = n - c$. 
The topological charge of the daughter soliton 
is characterized by $\pi_{n-1}$ in the bulk 
if it is a defect or gauge structure, 
or by $\pi_n$ if it is a texture.
It should be unchanged 
at A and B. 
In the world-volume theory of the mother soliton, 
the topological charge is characterized by $\pi_{m-1}$ if it is a defect, 
or by $\pi_m$ if it is a texture.

All previously known examples are listed in the next section as a review.
One of typical examples is given by confined monopoles 
(the second in a summary in the next section): 
In the Higgs phase, magnetic fluxes emanating from the monopole
are squeezed into flux tubes. 
If they are squeezed into two flux tubes, stable configuration 
of a monopole attached by two vortices from the both sides. 
In a certain situation, those vortices are 
non-Abelian vortices.
These two vortices can be regarded as a single non-Abelian vortex 
with non-Abelian magnetic fluxes directed in two opposite directions.
In this case, 
the monopoles can be represented by 
a kink in 
the effective world-sheet theory of 
a single 
non-Abelian vortex, which is the ${\mathbb C}P^{N-1}$ model \cite{Tong:2003pz}. 
This vortex monopole configuration 
has important applications to 
the correspondence  quantum effects in the two dimensional ${\mathbb C}P^{N-1}$ model and dimensional gauge theory \cite{Shifman:2004dr,*Hanany:2004ea}.

\begin{figure}
\begin{center}
\includegraphics[width=0.8\linewidth,keepaspectratio]{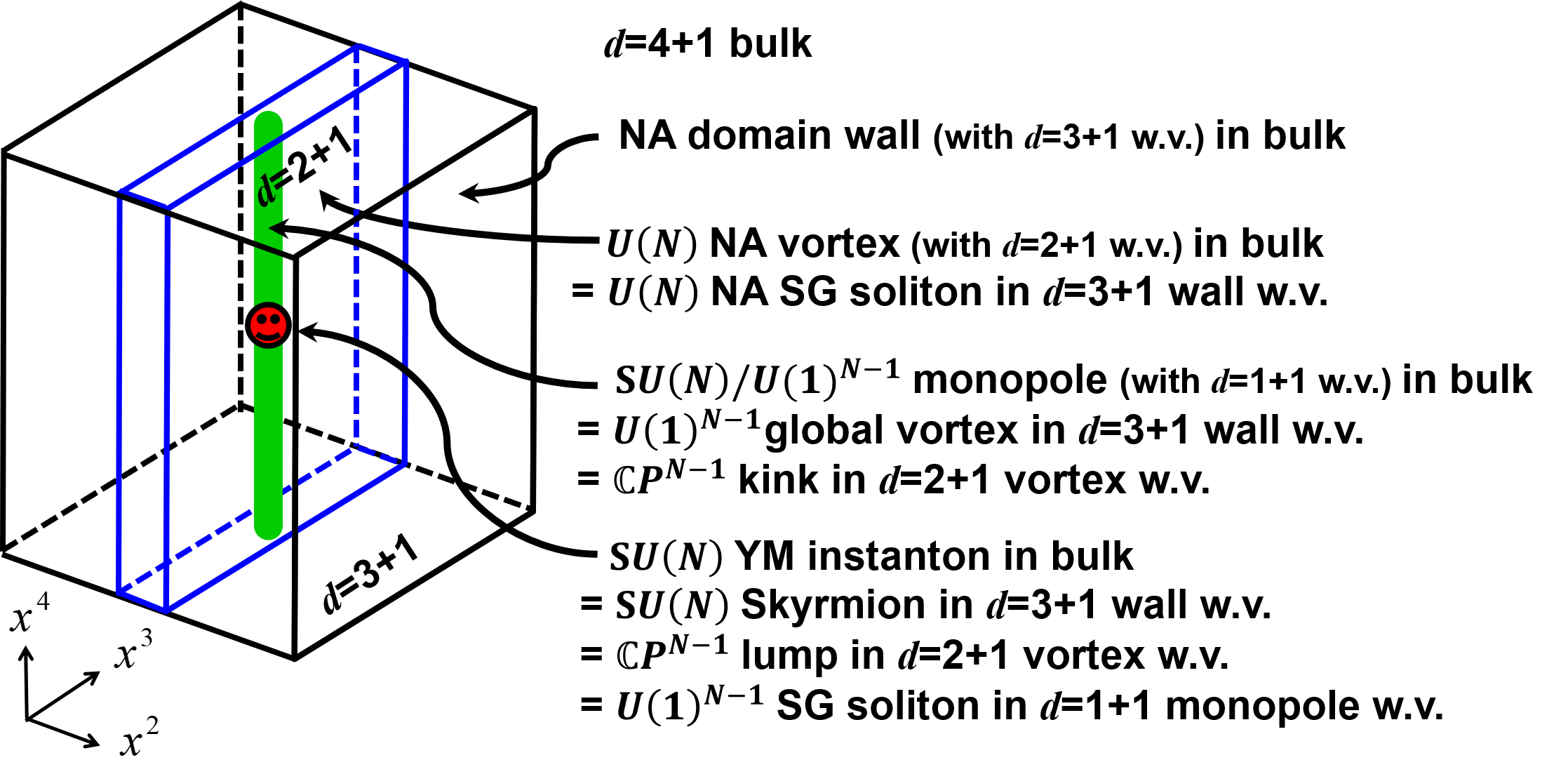}
\caption{A schematic picture of the configuration of 
the wall-vortex-monopole-instanton in $d=4+1$.
The codimensional direction $x^1$ of the wall is not shown.
The black box, blue sheet, green rod, and red circle 
denote the non-Abelian domain wall, non-Abelian vortex, magnetic monopole, and 
Yang-Mills instanton, respectively. 
YM, NA, SG and w.v. denote Yang-Mills, non-Abelian, sine-Gordon and world-volume, respectively.
\label{fig:wvmi}
}
\end{center}
\end{figure}

In this paper, as a configuration including all the composite solitons in the form of solitons within a soliton summarized in the next section, 
we construct  
a configuration consisting of four different solitons: 
a wall-vortex-monopole-instanton schematically drawn in Fig.~\ref{fig:wvmi}, 
which is the most general and unique configuration. 
Relations among topological solitons deduced from this configuration are summarized in Table \ref{table:relations}. 
Further hidden relations which can be obtained from 
Table \ref{table:relations} are summarized in 
Tables~\ref{table:relations2}
and \ref{table:relations3} (derivations are explained 
in subsequent sections).
We conjecture all possible relations among topological solitons 
can be obtained from this configuration. 
In our setup, Yang-Mills instantons, magnetic monopoles, non-Abelian vortices, 
and non-Abelian domain walls 
are primary solitons --- 
those who can live in the bulk. 
The others, 
Skyrmions, baby Skyrmions (lumps), 
(non-Abelian) sine-Gordon solitons, 
${\mathbb C}P^{N-1}$ lumps,  ${\mathbb C}P^{N-1}$ domain walls, 
and $U(1)^{N-1}$ global vortices   
are all secondary solitons --- those who can live only as daughter solitons inside a mother soliton. 
All relations among the secondary solitons can be deduced 
from Table  \ref{table:relations} as discussed in subsequent sections.
Throughout the paper, we use the term ``Yang-Mills instantons'' 
as codimensions four-objects in four Euclidean space 
in $d=4+1$ space-time.

\begin{table}
\begin{tabular}{c|c|ccc}
           &               &           & {\it Daughters}  &              \\
{\it Mothers} &  Moduli    &  $U(N)$ NA vortex &  $\frac{SU(N)}{U(1)^{N-1}}$ monopole & $SU(N)$ YM instanton \\ \hline
$U(N)$ NA wall      &    $U(N)$   & $U(N)$ NA  SG soliton  & \quad $U(1)^{N-1}$ global vortex  \quad     & $SU(N)$ Skyrmion \\
$U(N)$ NA vortex   &    ${\mathbb C}P^{N-1}$ &  --     &  
${\mathbb C}P^{N-1}$ wall  & ${\mathbb C}P^{N-1}$ lump \\
$\frac{SU(N)}{U(1)^{N-1}}$ monopole  & $U(1)$   &   --      &      --          & SG soliton
\end{tabular}
\caption{Relations among all kinds of solitons. 
Mother solitons can host daughter solitons inside them. 
The rightmost colum: An $SU(N)$ Yang-Mills instanton becomes 
an $SU(N)$ Skyrmion, 
${\mathbb C}P^{N-1}$ lump, 
and ($U(1)^{N-1}$ coupled) sine-Gordon soliton inside 
a non-Abelian domain wall, 
non-Abelian vortex 
and monopole, respectively. 
The middle column: An $SU(N)$ monopole becomes 
$U(1)^{N-1}$ global vortex 
and ${\mathbb C}P^{N-1}$ kink inside 
a non-Abelian domain wall and
non-Abelian vortex, respectively.
The leftmost column:
 A non-Abelian vortex becomes a non-Abelian sine-Gordon soliton 
 inside a non-Abelian vortex.
  \label{table:relations}
 }
\begin{tabular}{c|c|cc}
           &                          & {\it Daughters}  &              \\
{\it Mothers} &  Moduli    & $U(1)^{N-1}$ global vortex  & $SU(N)$ Skyrmion   \\ \hline
$U(N)$ NA SG soliton  &  ${\mathbb C}P^{N-1}$ & 
${\mathbb C}P^{N-1}$ wall & ${\mathbb C}P^{N-1}$ lump \\
$U(1)^{N-1}$ global vortex   & $U(1)$
& -- &
SG soliton    
   \\
\end{tabular}
\caption{Hidden relations (1) among solitons deduced from Table \ref{table:relations}.
  \label{table:relations2}
 }
\begin{tabular}{c|c|cc}
           &                          & {\it Daughter}  &              \\
{\it Mother} &  Moduli    &    ${\mathbb C}P^{N-1}$ lump (baby skyrmion)\\ \hline
${\mathbb C}P^{N-1}$ wall  & $U(1)$ & 
SG soliton 
\end{tabular}
\caption{Hidden relation (2) among solitons deduced from Table \ref{table:relations}.
  \label{table:relations3}
 }
\end{table}

This paper is organized as follows. 
In Sec.~\ref{sec:soliton-within-soliton}, we summarize all known composite solitons in the form of solitons within a soliton. 
In Sec.~\ref{sec:model}, we present our model, 
$U(N)$ gauge theory coupled with  Higgs  scalar fields.
In Sec.~\ref{sec:wvmi}, 
we construct the wall-vortex-monopole-instanton configuration 
in the effective field theory technique.
In Sec.~\ref{sec:relations}, we  obtain various secondary 
relations deduced from the configuration.
Section \ref{sec:summary} is devoted to a summary 
and discussion. 
In Appendices, we summarize solitons within a soliton which we 
do not use in this paper.  
In Appendix \ref{sec:same-dim}, we summarize composite solitons 
in which mother and daughter solitons have the same dimensions.
In Appendix \ref{sec:tbc}, we summarize composite solitons 
in compactified spaces ${\mathbb R}^{d-1} \times S^1$ 
with twisted boundary conditions.

\newpage

~

\newpage
\section{Solitons within a soliton: a review}\label{sec:soliton-within-soliton}

There are already several known such composite solitons 
in the form of solitons within a soliton 
 in the literature,
deducing relations among topological solitons.
In this section, let us summarize all known composite solitons 
in the form of solitons within a soliton.
In the first subsection, we summarize 
composite solitons consisting of two different solitons, 
which we call composite solitons of 2-generations.
Then, in the second subsection, we summarize 
composite solitons of 3-generations.

\subsection{2-generations (mother and daughter)}

We use a rule for names of composite solitons as follows
-- Mother daughter.
\begin{enumerate}

\item 
{\bf Domain-wall baby-Skyrmions(lumps), or domain-wall vortices}:

The simplest example of composite solitons is 
a domain-wall baby-Skyrmion.
\begin{enumerate}
\item
${\mathbb C}P^1$ lumps (baby Skyrmions, or semilocal vortices) are sine-Gordon solitons inside a ${\mathbb C}P^1$ wall 
\cite{Nitta:2012xq,*Kobayashi:2013ju} 
(see also \cite{Kudryavtsev:1997nw,*Auzzi:2006ju,*Jennings:2013aea,*Bychkov:2016cwc}).
Theoretically, this setup provides a physical proof for
the lower dimensional analogue of the Atiyah-Manton construction
\cite{Sutcliffe:1992he,*Stratopoulos:1992hq}. 
Physically, this configuration is realized in two condensed matter systems. 
The first is ``domain-wall Skyrmions'' in chiral magnets \cite{PhysRevB.99.184412,*PhysRevB.102.094402,*Nagase:2020imn,*Yang:2021} 
(see also \cite{Kim:2017lsi}). 
(In the condensed matter community, baby Skyrmions are simply 
called Skyrmions.)
See also Ref.~\cite{Hongo:2019nfr} 
for domain-wall instantons in $d=1+1$ chiral magnets. 
The second is Josephson vortices (also called fluxons) in a Josephson junction 
\cite{USTINOV1998315,Ustinov2015}. 
A Josephson junction can be regarded as 
a heavy tension limit of a domain wall 
in a $U(1)$ gauge theory coupled with two complex scalar fields
\cite{Nitta:2012xq,*Kobayashi:2013ju}.
The domain wall separates 
two vacua with only one of the complex scalar fields developing  
a VEV, and thus the wall is an insulator region of 
the Josephson junction.
Then, vortices in the bulk are absorbed into the wall 
to become Josephson vortices (fluxons), whose dynamics  
can be described as sine-Gordon solitons 
in the domain wall effective theory.

\item
${\mathbb C}P^{N-1}$ lumps are 
$U(1)^{N-1}$ coupled sine-Gordon solitons inside 
${\mathbb C}P^{N-1}$ walls \cite{Fujimori:2016tmw}.

The aforementioned Josephson junction of two superconductors 
can be generalized to multi-layered $N$ parallel Josephson junctions.
This case can be described by $N-1$ parallel domain walls
in the ${\mathbb C}P^{N-1}$ model  \cite{Fujimori:2016tmw}.
A single Josephson vortex (fluxson) can be described in each wall, 
while 
the interaction between Josephson vortices 
in different domain walls can be also studied 
\cite{Fujimori:2016tmw}.

\item
Non-Abelian vortices are non-Abelian sine-Gordon solitons 
inside a non-Abelian domain wall 
\cite{Nitta:2015mma}. 
A non-Abelian domain wall 
\cite{Shifman:2003uh,Eto:2005cc,Eto:2008dm} 
can describe 
a non-Abelian Josephson junction 
(of two color superconductors) 
once a Josephson term is introduced.
The effective theory of the non-Abelian domain wall is then 
a $U(N)$ chiral Lagrangian with a pion mass
or non-Abelian sine-Gordon model  
\cite{Nitta:2015mma,Nitta:2015mxa}. 
Then, non-Abelian vortices 
\cite{Hanany:2003hp,*Auzzi:2003fs} absorbed into 
the domain wall  
become non-Abelian Josephson vortices 
described as a non-Abelian sine-Gordon soliton \cite{Nitta:2014rxa,Eto:2015uqa}.

\end{enumerate}

Similar (but different) configurations that cannot be regarded as solitons within a soliton can be found.\footnote{
Similar configurations can be found as
axion domain wall(s) ending on an axion string 
\cite{Vilenkin:1982ks,*Kawasaki:2013ae},
axial (chiral) domain wall(s) ending on 
an axial (chiral) vortex in QCD 
\cite{Balachandran:2001qn,Balachandran:2002je,*Eto:2013bxa,Eto:2013hoa}, 
and wall-vortex composites in 
GUT \cite{Kibble:1982dd,*Everett:1982nm},
2HDMs  
\cite{Dvali:1993sg,*Eto:2018hhg,*Eto:2018tnk,*Eto:2021dca},  
the Gerogi-Machacek model 
\cite{Chatterjee:2018znk},
and 
supersymmetric theories \cite{Ritz:2004mp,*Bolognesi:2007bh}. 
These cases are not solitons within a soliton 
except for some special cases 
in which a vortex is attached by two domain walls of the same tension.
The wall decays quantum mechanically \cite{Preskill:1992ck}.
}

\item
{\bf Vortex monopoles}:

The second simplest example of composite solitons is 
a vortex monopole.
\begin{enumerate}
\item
Monopoles are ${\mathbb C}P^{N-1}$ kinks inside a non-Abelian vortex (confined monopoles) 
\cite{Tong:2003pz}. 
This vortex monopole configuration 
(as well as vortex instanton configuration given below)
has important applications to 
the correspondence of quantum effects in the two dimensional ${\mathbb C}P^{N-1}$ model and four dimensional gauge theory \cite{Shifman:2004dr,*Hanany:2004ea}.

\item
Non-Abelian monopoles on multiple non-Abelian vortices 
can be also realized \cite{Nitta:2010nd}. 
Mathematically, this provides a physical realization  
of the Donaldson's rational map \cite{Donaldson:1984ugy}.

\item
This correspondence can be generalized to
$SO, USp$ groups \cite{Eto:2011cv}.

\end{enumerate}
These can be obtained from vortex instantons given below 
by the Scherk-Schwarz dimensional reduction 
\cite{Scherk:1979zr}
or an $S^1$ compactification with a twisted boundary condition 
\cite{Eto:2004rz}.
There are similar (but different) configurations that 
cannot be regarded as soliton within a soliton.\footnote{
As similar configurations, there are 
vortices ending on monopoles  
in QCD \cite{Nambu:1974zg,*Mandelstam:1974vf,*Mandelstam:1974pi}, GUTs \cite{Langacker:1980kd}, 
supersymmetric QCD 
\cite{Auzzi:2003em,*Auzzi:2004yg,*Eto:2006dx,*Cipriani:2011xp,*Chatterjee:2014rqa}, 
and high density QCD \cite{Gorsky:2011hd,*Eto:2011mk}.
A Nambu monopole in the Standard Model 
\cite{Nambu:1977ag,Achucarro:1999it,Eto:2012kb} 
is attached by a single $Z$-string, similar to the above.
On the other hand, a Nambu monopole in 2HDMs \cite{Eto:2019hhf,*Eto:2020hjb,*Eto:2020opf} 
is attached by two non-Abelian $Z$-strings of the same tension, 
similar to vortex monopoles in the topic of this paper. 
However, it cannot be regarded as a kink on a single $Z$-string 
since magnetic fluxes spread out spherically from the monopole.
See further Refs.~\cite{Hindmarsh:1985xc,*Kibble:2015twa,*Kleihaus:2003nj,*Kleihaus:2003xz,*Ng:2008mp} for similar objects. 
These cases are not solitons within a soliton 
except for some special cases 
in which a monopole is attached by vortex strings of the same tension.
}

\item
{\bf Domain-wall monopoles}:

Monopoles are $U(1)^{N-1}$ global vortices inside a non-Abelian domain wall \cite{Nitta:2015mxa}. 
 We call them Josephson monopoles. 
These can be obtained from domain-wall instantons given below 
by the Scherk-Schwarz dimensional reduction 
\cite{Scherk:1979zr}
or an $S^1$ compactification with a twisted boundary condition.

\item
{\bf Vortex instantons}:

Yang-Mills instantons are ${\mathbb C}P^{N-1}$ lumps inside a non-Abelian vortex 
\cite{Hanany:2004ea,Eto:2004rz,Fujimori:2008ee}. 
Fractional instantons and calorons are 
transformed to vortex monopoles   with the twisted boundary condition  
\cite{Eto:2004rz}.
This composite soliton can be applied for
correspondence between the vortex counting 
and instanton counting
\cite{Fujimori:2015zaa,*Fujimori:2012ab}.

\item
{\bf Domain-wall instantons}:

Yang-Mills instantons are Skyrmions inside a non-Abelian domain wall 
\cite{Eto:2005cc} (see also \cite{Nitta:2013vaa,Nitta:2015mxa}).
We also call these instantons Josephson instantons. 
Originally, the term ``domain-wall Skyrmion'' was used for this configuration, 
but in this paper we call it a ``domain-wall instanton'' 
because of the rule of names (mother-daughter).
This configuration provides a proof of 
the Atiyah-Manton construction of Skyrmions \cite{Atiyah:1989dq,*Atiyah:1992if}.
The situation is also similar to holographic QCD \cite{Hata:2007mb}.

\item
{\bf Vortex Skyrmions}:

Skyrmions are sine-Gordon solitons inside a global vortex 
in a Skyrme model with a twisted mass term \cite{Gudnason:2016yix}.
Similar configurations can be found in a BEC-Skyrme model 
(Skyrme model with a BEC motivated potential term)
\cite{Gudnason:2014hsa,*Gudnason:2014gla}.

\item
{\bf Domain-wall Skyrmions and sine-Gordon Skyrmions} :

The final examples are 
domain-wall Skyrmions and sine-Gordon Skyrmions.
\begin{enumerate}

\item
Skyrmions are baby Skyrmions (lumps) inside a ``non-Abelian'' $S^2$ domain wall 
 \cite{Nitta:2012wi,*Gudnason:2014nba} (see also Refs.~\cite{Kudryavtsev:1999zm,*Gudnason:2013qba}).  
 This configuration can be generalized to
 higher dimensions (called matryoshka Skyrmions) \cite{Nitta:2012rq}.
 Similar configurations can be found in a BEC-Skyrme model \cite{Gudnason:2018oyx}.
This configuration provides 
a translational (or Donaldson's type) rational map ansatz of $SU(2)$ Skyrmions \cite{Houghton:1997kg}. 
Physically, domain wall Skyrmions were discussed in 
a chiral soliton lattice in QCD \cite{Chen:2021vou}.

\item
Skyrmions are ${\mathbb C}P^{N-1}$ lumps inside a non-Abelian sine-Gordon soliton \cite{Eto:2015uqa}.
This configuration provides a translational (or Donaldson's type) rational map ansatz of $SU(N)$ Skyrmions \cite{Ioannidou:1999mf}.

\end{enumerate}

\item {\bf Lump-string Hopfions}:

Hopfions are sine-Gordon solitons inside a ${\mathbb C}P^1$ lump-string.

\end{enumerate}
These are all known examples of composite solitons consisting of two generations.
Some cases are BPS states preserving a fraction of supersymmetries if one embeds the theories to supersymmetric theories 
\cite{Eto:2005sw}.

\subsection{3-generations (grandmother, mother and daughter)}

In addition, there are a few examples of three generations, 
that is, composite solitons of three different dimensions.
We use a rule for names of composite solitons as grandmother-mother-daughter. 
\begin{enumerate}
\item
{\bf Vortex-monopole-instanton}: 

Yang-Mills instantons inside a monopole inside a non-Abelian vortex 
\cite{Nitta:2013cn} 
(see also Refs.~\cite{Nitta:2012mg, Nitta:2013vaa}).
Instantons are sine-Gordon solitons inside the monopole 
and lumps inside the vortex at the same time, while the monopole is a kink inside the vortex.

\item
{\bf Wall-vortex-instanton}:

Yang-Mills instantons inside a vortex inside a non-Abelian domain wall 
 \cite{Nitta:2015mxa}.
Instantons are lumps inside a vortex and Skyrmions inside the domain wall at the same time, while the vortex is a non-Abelian sine-Gordon soliton inside the domain wall.

\item
{\bf Wall-monopole-instanton}: 

Yang-Mills instantons inside a monopole inside a non-Abelian domain wall 
\cite{Nitta:2015mxa}.
Instantons are sine-Gordon solitons inside a monopole and Skyrmions inside the domain wall at the same time, while the monopole is a kink inside the vortex. 
This configuration can be obtained as the Scherk-Schwarz dimensional reduction \cite{Scherk:1979zr}
of the wall-vortex-instanton.

\end{enumerate}

As already mentioned, in the following sections, 
we construct  
a configuration of four generations 
as a configuration including all the aforementioned composite solitons in the form of solitons within a soliton of two or three generations: 
a wall-vortex-monopole-instanton schematically drawn in Fig.~\ref{fig:wvmi}.

\section{The model: non-Abelian gauge theory in the Higgs phase \label{sec:model}}
The theory that we consider is 
a $U(N)$ gauge theory coupled with Higgs scalar fields 
in the Higgs phase 
in $d=4+1$ dimensions 
with the following matter contents: 
a $U(N)$ gauge field $A_{\mu}(x)$,  
two $N$ by $N$ charged complex scalar fields 
$H(x) = (H_1(x),H_2(x))$,  
and a real neutral adjoint  
$N$ by $N$ scalar field $\Sigma(x)$.  
The $U(N)$ gauge (color) symmetry acts on fields as
\beq
\,
 A_{\mu} \to g A_{\mu} g^{-1} + i g \del_{\mu} g^{-1},  \quad
 H \to g H , \quad 
 \Sigma \to g \Sigma g^{-1}, \quad 
\, g \in U(N)_{\rm C}.
\eeq
The Lagrangian is given as follows:
\beq
&& {\cal L} \,=\, -\1{4 g^2} \tr F_{\mu\nu}F^{\mu\nu} 
 + \1{g^2} \tr (D_{\mu} \Sigma)^2\, 
 + \tr |D_{\mu} H|^2 + {\cal L}_J  - V 
 \label{eq:Lag}
\eeq
where $V$ is the potential term
\beq
&& V \,=\, {g^2 \over 4} \tr ( H H^\dagger -v^2 {\bf 1}_N)^2 
 \,+ \tr |\Sigma H - H M|^2 ,
\eeq
and $D_{\mu}$ is the covariant derivative, 
given by  
$D_{\mu} H = \del_{\mu} H - i A_{\mu } H$ 
and 
$D_{\mu} \Sigma = \del_{\mu} \Sigma - i [A_{\mu},\Sigma]$, 
$g$ is the gauge coupling constant that we take 
common for 
the $U(1)$ and $SU(N)$ factors of $U(N)$, 
$v$ is a real constant representing the 
vacuum expectation value of $H$, and   
$M$ is a $2N$ by $2N$ mass matrix for $H$ 
given below.
Apart from the Josephson term $ {\cal L}_J$, 
the model is a truncation of the bosonic part of 
${\cal N}=2$ supersymmetric 
theory (with eight supercharges) 
in $d=4+1$ \cite{Eto:2006pg}. 

In the massless case $M =0$, 
the flavor symmetry is the maximum $SU(2N)$.
This is explicitly broken by 
the mass matrix $M$ that we take 
\beq 
 M={\rm diag.}(m {\bf 1}_N + \Delta M, 
- m {\bf 1}_N - \Delta M) , \quad
\Delta M = {\rm diag.} (m_1,m_2,\cdots,m_N)
\label{eq:mass}
\eeq
with a real mass $m$ 
and real mass shifts $m_a$ much smaller than $m$:
$m_a \ll m$.
For $m \neq 0$ with $\Delta M=0$, the flavor symmetry  is
 $SU(N)_{\rm L} \times  SU(N)_{\rm R} \times U(1)_{\rm L-R}$, given by 
\beq
 H_1 \to H_1 U_{\rm L} e^{+i \alpha}, \quad 
 H_2 \to H_2 U_{\rm R} e^{-i \alpha},  \quad 
 U_{\rm L,R} \in SU(N)_{\rm L,R}, \quad 
 e^{i\alpha} \in  U(1)_{\rm L-R} ,
\label{eq:flavor}
\eeq
while for $\Delta M \neq 0$ with 
non-degenerate mass perturbation 
$m_a \neq m_b$ for $a \neq b$, 
the flavor symmetry is further explicitly broken to 
$U(1)^{N-1}_{\rm L} \times U(1)^{N-1}_{\rm R}\times U(1)_{\rm L-R}$.
In this paper, we consider the non-degenerate masses.
Without loss of generality, we can assume $m_r > m_{r+1}$.

In the Lagrangian in Eq.~(\ref{eq:Lag}),  
${\cal L}_J$ consists of scalar couplings that we call 
the Josephson interactions 
\beq
&&
 {\cal L}_{J} \,=\,  {\cal L}_{J,1} +  {\cal L}_{J,2},\\
&& {\cal L}_{J,1} \,=\, -
 \gamma \tr (H_1^\dagger H_2  + H_2^\dagger H_1) \,\,
\label{eq:Josephson}\\
&& {\cal L}_{J,2}  =  - \sum_{r=1}^{N-1} {\beta_r^2 \over v^2} 
 [\tr (H_1 X_r H_2^\dagger ) + \tr (H_2 X_r H_1^\dagger )] \label{eq:deform}
\eeq
where $X_r = X_r^\dagger$ ($r=1,\cdots,N-1$) 
are elements of $SU(N)$ algebra 
into which $\sigma_1$ is embedded as a diagonal submatrix.
Here, 
${\cal L}_{J,1}$ gives a Josephson interaction of 
two color superconductors separated by a domain wall, 
given below.

The vacuum structures of the model are as follows. 
In the massless case $m=0$, $\Delta M =0$, 
$\gamma=0$, $\beta_r=0$, 
the vacuum can be taken without the lost of generality as
\beq
\, H 
\,= \,
\left(
 v {\bf 1}_N ,{\bf 0}_N 
\right) ,\,
\quad
\quad \Sigma ={\bf 0}_N\,\,
\eeq 
by using the $SU(2N)$ flavor symmetry.
The unbroken symmetry is 
$SU(N)_{\rm C+L} \times SU(N)_{\rm R} \times U(1)$, 
in which the factor $SU(N)_{\rm C+L}$ is 
the color-flavor locked (global) symmetry.
The moduli space of vacua is  
the complex Grassmann manifold 
\cite{Higashijima:1999ki} 
\beq  
\,\, Gr_{2N,N}
 \,\simeq\, {SU(2N) \over SU(N)\times SU(N)\times U(1)}.\,
\label{eq:Grassmann}
\eeq 

In the massive case, $m \neq 0$ but still $\Delta M =0$, 
the above vacua are split into the following two disjoint vacua  
\beq
&& H \,=\, 
\left(
 v {\bf 1}_N ,{\bf 0}_N 
\right) , \quad
\Sigma = + m {\bf 1}_N:  
 \quad SU(N)_{\rm C+L} ,\,
\non
&& H \,=\, 
\left(
  {\bf 0}_N , v{\bf 1}_N 
\right) ,  \quad \Sigma = - m {\bf 1}_N:
\quad SU(N)_{\rm C+R} \,
 \label{eq:vac}
\eeq 
with the unbroken color-flavor locked (global) symmetries
$g=U_{\rm L}$ and $g = U_{\rm R}$, respectively.
These vacua are
color-flavor locked vacua  
that can be interpreted as 
non-Abelian color superconductors.  
With the non-degenerate mass deformation 
$\Delta M \neq 0$, each vacuum 
in Eq.~(\ref{eq:vac}) is shifted to
\beq
&& H = 
\left(
 v' {\bf 1}_N ,{\bf 0}_N 
\right) , \quad
\Sigma =  + m {\bf 1}_N + \Delta M:  
 \quad U(1)^{N-1}_{\rm C+L} ,
\non
&& H = 
\left(
  {\bf 0}_N , v' {\bf 1}_N 
\right) ,  
\quad \Sigma = - m {\bf 1}_N - \Delta M:
\quad U(1)^{N-1}_{\rm C+R},
 \label{eq:vac2}
\eeq 
where $v'$ is shifted from $v$.

In the following sections, 
we often work in the strong  coupling 
(nonlinear sigma model) limit 
$g \to \infty$ 
for explicit calculations.  
In this limit, we have the constraints 
\beq
H H^\dagger = v^2 {\bf 1}_N,
\quad
\Sigma 
= v^{-2} H M H^\dagger,
\quad
A_{\mu} 
= {i\over 2} v^{-2} [H \del_{\mu} H^\dagger -  (\del_{\mu}H)  H^\dagger], \, \label{eq:Amu}
\eeq
and the model is reduced 
to the Grassmann sigma model 
with the target space given in Eq.~(\ref{eq:Grassmann})
together with a potential term, known as the massive (twisted-mass deformed) Grassmann sigma model \cite{Arai:2003tc}.
In this limit, vortices reduce to Grassmann sigma model lumps.

In this paper, we consider the following 
hierarchical symmetry breakings:
\beq
 && m \quad \gg \quad \gamma \quad \gg \quad \Delta m_r \quad \gg \quad  \beta_r   
 \label{eq:hierarchy}  \\
 && \mbox{wall \quad vortex  \quad monopole  \quad instanton}\nonumber
\eeq
with $\Delta m_r \equiv m_r -m_{r+1}$ ($r=1,2,\cdots, N-1$).
The second line denotes topological solitons 
that form when turning on the corresponding parameters. 
In the following, we turn on these parameters 
from the left to right gradually.

\section{Wall-Vortex-Monopole-Instanton
\label{sec:wvmi}}

In this section, we construct the wall-vortex-monopole-instanton 
configuration 
in the moduli approximation, 
by gradually turning of 
hierarchical parameters in Eq.~(\ref{eq:hierarchy}).
In each subsection we construct 
the wall, vortex, monopole, and instanton
with turning on
 $m$, $\gamma$, $\Delta m_r$, and  $\beta_r$ 
  in Eq.~(\ref{eq:hierarchy}), 
  respectively.

\subsection{The first hierarchy: non-Abelian domain wall}

We first turn on the mass $m$ in the hierarchy in Eq.~(\ref{eq:hierarchy}).
In the sigma model limit,
a non-Abelian domain wall solution  
interpolating between the two vacua 
in Eq.~(\ref{eq:vac}) 
perpendicular to the coordinate $x^1$
can be given by 
\cite{Isozumi:2004jc,Shifman:2003uh,
Eto:2005cc,Eto:2008dm}
\beq 
&& H = \,H_{\rm wall} (x^1)
\,=  {v\over \sqrt {1+ e^{\mp 2 m (x^1-X^1) }}}
      \left({\bf 1}_N, e^{\mp m (x^1-X^1) }U \right), 
  \label{eq:wall-sol}
\eeq 
with $\Sigma$ and $A_1$ in
Eq.~(\ref{eq:Amu}).
The width of the wall is $m^{-1}$. 
Here, 
 $X^1$ is the position (translational modulus) 
 of the domain wall in 
 the coordinate $x^1$ and 
$U$ contains group-valued moduli 
$U \in U(N)$: 
\beq 
\, (X^1,U) \in  {\cal M}_{\rm wall} \,\simeq \, {\mathbb R} \times U(N).
\, \label{eq:wall-moduli}
\eeq 

The effective theory of the non-Abelian 
domain wall can be 
constructed by using 
the moduli approximation \cite{Manton:1981mp,Eto:2006uw};
First, we promote the moduli parameters $X^1$ and $U$ to moduli fields 
 $X^1(x^i)$ and $U(x^i)$, respectively ($i=0,2,3,4$) 
on the world-volume of the domain wall,  
and then perform integration over the codimension $x^1$. 
We thus obtain the effective theory given by \cite{Shifman:2003uh,Eto:2005cc,Eto:2008dm}:
\beq
 \,{\cal L}_{\rm wall} 
 &=&  \int dx^1 {\cal L} (H=H_{\rm wall}(x^1; X^1(x^i),U(x^i)))\non
 &=&
{v^2 \over 2m} \del_i X^1 \del^i X^1 
-  f_{\pi}^2 
\tr \left(U^\dagger \del_{i} U
            U^\dagger \del^{i} U \right) ,
\,
\quad  f_{\pi}^2 \equiv {v^2 \over 4m} ,
\label{eq:eff0}
\eeq
which is a $U(N)$ chiral Lagrangian, 
or principal chiral model.
If we calculate the next leading order of the derivative expansion 
of the effective theory, we would obtain the Skyrme term 
\cite{Eto:2005cc}.

\subsection{The second hierarchy: non-Abelian vortex 
trapped inside the wall}

Now 
we turn on the Josephson interaction $\gamma$ 
as the second largest parameter in the hierarchy in Eq.~(\ref{eq:hierarchy}), 
so that the domain wall becomes the Josephson junction. 
Let us evaluate the effect of $\gamma$ perturbatively  
in the domain wall effective theory in Eq.~(\ref{eq:eff0}) 
provided that $\gamma$ is small enough not to 
affect the domain wall solution at the leading order. 
We thus find  
the pion mass term  
\cite{Nitta:2015mma}
\beq
\Delta {\cal L}_{{\rm wall},J}
&=& \int dx^1 {\cal L}_{J,1} (H=H_{\rm wall}(x^1; X^1(x^i),U(x^i)))\non
&=& - m'^2  (\tr U + \tr U^\dagger) ,\quad
\label{eq:eff-Josephson}
m'^2 \equiv  {\pi \gamma \over 2m}.
\eeq
This potential term lifts 
the $U(N)$ vacuum manifold,  
leaving the unique vacuum
$U = {\bf 1}_N$
as the case of the usual chiral Lagrangian.

When the non-Abelian vortex is placed parallel to 
the non-Abelian Josephson junction (domain wall), 
it is absorbed into the junction 
to minimize the total energy. 
The resulting configuration can be described as 
a non-Abelian sine-Gordon soliton
in the $U(N)$ chiral Lagrangian in Eq.~(\ref{eq:eff0}) 
with the mass term in Eq.~(\ref{eq:eff-Josephson}). 
A non-Abelian sine-Gordon soliton 
(perpendicular to the $x^2$ coordinate) is given by  
\cite{Nitta:2014rxa,Eto:2015uqa}:
\beq
&& \,U \,=\, U_{\rm vortex}(x^2)=V {\rm diag}\, (u(x^2),1,\cdots,1) V^\dagger
= {\bf 1}_N + (u-1) \phi \phi^\dagger
\non 
&& \,u(x^2) \,=\, \exp i \theta_{\rm SG}(x^2-X^2) 
= \exp \left(4 i \, \arctan \exp [m''  (x^2- X^2)] \right) ,\,
\label{eq:U(1)-one-kink}
\eeq
with $m''^2 = \frac{m'^2}{f_\pi^2} =\frac{2\pi \gamma}{v^2}$.
The width of the soliton is $m''^{-1} \sim v/\sqrt{\gamma}$, 
and the tension of the soliton is  
$T_{\rm SG} = 8m''$. 
Here $X^2$ is translational modulus in the coordinate 
$x^2$, and $V$ are orientational moduli taking a value in 
${\mathbb C}P^{N-1} \simeq {SU(N) \over SU(N-1) \times U(1)}$, 
or more precisely
$\phi \in {\mathbb C}^N$ ($\phi^\dagger \phi =1$) are  
homogeneous coordinates of ${\mathbb C}P^{N-1}$.
Therefore, the moduli of 
the non-Abelian sine-Gordon soliton are 
\beq
 {\cal M}_{\rm vortex}  \simeq  {\mathbb R}\times {\mathbb C}P^{N-1} 
\eeq
that coincide with the moduli of the non-Abelian vortex 
in the bulk, 
except for 
one translation modulus fixed to be the position $X^1$ of the wall.
It was also shown in Ref.~\cite{Nitta:2015mma}
from the flux matching that this is precisely a non-Abelian vortex.

The effective theory of the sine-Gordon soliton with the world-volume 
$x^{\alpha}$ ($\alpha=0,3,4,5$) can 
be also obtained by the moduli approximation \cite{Eto:2015uqa}:
\beq
{\cal L}_{\rm vortex} &=& 
\int dx^2 {\cal L}_{\rm wall} (U = U_{\rm vortex } (x^2;X^2(x^\alpha), \phi (x^{\alpha}) )) 
\non
&=&
C_X
\p_\alpha X^2 \p^\alpha X^2 
+ C_{\phi} 
\left[
\p_\alpha \phi^\dagger \p^\alpha \phi + (\phi^\dagger\p_\alpha \phi)(\phi^\dagger\p^\alpha \phi)
\right]  \label{eq:vortex-eff}
\eeq
with the constants 
$C_X =  \frac{f_\pi^2 T_{\rm SG}}{2} = \sqrt{2\pi} \frac{v{\sqrt \gamma}}{m}$
and 
$C_{\phi} = \frac{f_\pi^2 T_{\rm SG}}{m''^2} = \sqrt{2\over \pi}{v^3 \over m \sqrt \gamma} $.
This is the ${\mathbb C}P^{N-1}$ model 
written in terms of the homogeneous coordinates $\phi$.
The four derivative correction to the vortex effective theory
 was obtained for a non-Abelian vortex in the bulk 
 \cite{Eto:2012qda}, 
 but we do not need it in our study.

\subsection{The third hierarchy: monopole trapped inside the vortex inside the wall}

Now let us turn on $\Delta m_r$ (in $\Delta M$)
 in the hierarchy in Eq.~(\ref{eq:hierarchy}).
First, $\Delta M$ induces a twisted mass in the $U(N)$ chiral Lagrangian
as the domain wall effective theory in Eq.~(\ref{eq:eff0}).
This potential can be obtained by the Scherk-Schwarz dimensional reduction \cite{Nitta:2015mxa}: 
\beq
V_{\rm wall} = 
   {v^2 \over 4m}  \tr ([ \Delta M, U ]^\dagger[ \Delta M, U ]),
\eeq
implying that diagonal $U$ has lower energy. 
The sine-Gordon soliton (vortex in the bulk) in Eq.~(\ref{eq:U(1)-one-kink}) should be either of $N$ diagonal embedding.

In fact,
the vortex effective theory 
in Eq.~(\ref{eq:vortex-eff})
is also modified by the twisted mass $\Delta M$ 
as \cite{Nitta:2015mxa}:
\beq
 V_{\rm vortex} 
&=&  C_{\phi}
\left[(\phi^\dagger \Delta M \phi)^2-\phi^\dagger (\Delta M)^2 \phi 
\right] . \label{eq:massive-CP}
\eeq
The Lagrangian in Eq.~(\ref{eq:vortex-eff}) 
with this potential term
is known as the massive ${\mathbb C}P^{N-1}$ model.
For non-degenerate mass deformation $\Delta M$,  
this potential admits $N$ discrete  
vacua
\beq
 \phi_a^T = v (0,\cdots,0,1,0,\cdots) ,  \quad a=1,\cdots,N
\eeq
where only the $a$-component is nonzero. 
These correspond to embedding of $u$ into diagonal elements in 
the solution $U$ in Eq.~(\ref{eq:U(1)-one-kink}).

For later use, it is often enough to consider 
a ${\mathbb C}P^1$ submanifold. 
The vortex effective theory of 
the $r$-th ${\mathbb C}P^1$ submanifold 
parametrized by a homogeneous coordinate $u$ 
in $\phi = \1{\sqrt{1+|u|^2}} (0, \cdots,0,1,u,0\cdots)$, 
where only $r$-th and $(r+1)$-th components are nonzero, can be written as
\beq
&& {\cal L}_{\rm vortex,{\mathbb C}P^1}
=  
C_X \p_\alpha X^2 \p^\alpha X^2 
+ C_{\phi} \left[ 
 {\partial_{\alpha} u^* \partial^{\alpha} u - \delta m_r^2 |u|^2 
  \over (1 + |u|^2)^2} \right] .\label{eq:vortex-th}
\eeq
with $\delta m_r \equiv m_{r+1} - m_r$.
The vacua in the vortex theory, $u=0$ (the north pole) 
and $u=\infty$ (south pole) of the target space 
${\mathbb C}P^1$,
correspond to embedding of $u$ to the upper-left 
and lower-right elements of the non-Abelian 
sine-Gordon solution $U$ 
in Eq.~(\ref{eq:U(1)-one-kink}), respectively.

The Lagrangian (\ref{eq:massive-CP})
admits $N-1$ multi-kink solutions 
\cite{Gauntlett:2000ib,*Tong:2002hi,Isozumi:2004jc,*Isozumi:2004va,*Isozumi:2004vg,*Eto:2004vy}, 
where 
the constituent kink connecting the 
$r$-th and $(r+1)$-th vacua has the mass 
$E_{{\rm kink},r}$ ($r=1,\cdots,N-1$) 
which is proportional to 
the mass of a monopole $E_{{\rm monopole},r}$ 
\cite{Tong:2003pz,Shifman:2004dr,*Hanany:2004ea,Nitta:2010nd}:
\beq 
 E_{{\rm kink},r} = C_{\phi} \delta m_r  ,\quad 
   E_{{\rm monopole},r} = {4\pi \over g^2} \delta m_r .
   \label{eq:monopole-energy}
\eeq
When the vortex goes to the bulk outside the wall where 
the vortex becomes BPS, 
the \kahler moduli $C_\phi$ in the vortex effective theory in Eq.~(\ref{eq:massive-CP}) is replaced by ${4\pi \over g^2}$.
Accordingly, 
the energy of kinks on the vortex becomes  
$E_{{\rm kink},r} =   E_{{\rm monopole},r} = {4\pi \over g^2} \delta m_r$. 
Then, one can confirm the kinks on the vortex represent the monopoles, and from the BPS properties, 
they carry correct monopole charges.
When the vortex gets back into the wall, 
the energy becomes the first in 
Eq.~(\ref{eq:monopole-energy}) 
different from the monopole energy in the bulk 
[the second in Eq.~(\ref{eq:monopole-energy})], 
but still carrying the same monopole charges.

A single monopole solution can be constructed 
by restricting ourselves to the $r$-th ${\mathbb C}P^1$ submanifold 
in Eq.~(\ref{eq:vortex-th}). 
It is a domain wall interpolating the two vacua 
$u=0$ and $u=\infty$ \cite{Abraham:1992vb} in the vortex effective theory (\ref{eq:vortex-th}).
We place it in the $x^3$-coordinate as
\beq
 u = u_{\rm monopole}(x^3) = e^{\mp \delta m_r (x^3-X^3) + i \ph} ,
\eeq
where $\mp$ represents a monopole and an anti-monopole 
with the width $1/\delta m_r$.
Here, $X^3$ and $\ph$ are moduli parameters 
representing the position in the $x^3$-coordinate and $U(1)$ phase 
of the (anti-)monopole. 
The moduli  of the monopole is then 
\beq
 {\cal M}_{\rm monopole}  \simeq {\mathbb R}\times U(1)
\eeq 
coinciding with the monopole moduli except for 
two translational moduli fixed to the positions 
$X^1$ and $X^2$
of the domain wall 
and the vortex.

Let us construct the effective theory of the single monopole-string 
by promoting the moduli $X^3$ and $\ph$ to fields 
 $X^3(x^m)$ and $\ph(x^m)$ ($m=0,4$) 
 \cite{Manton:1981mp,Eto:2006uw} 
on the monopole string  \cite{Nitta:2013cn}: 
\beq
 {\cal L}_{r-{\rm th \; monopole}} 
 &=& \int dx^3 {\cal L}_{{\rm vortex}, {\mathbb C}P^1} 
 (u = u_{\rm monopole} (x^3; X^3(x^m) ,\ph(x^m)) )  \non
&=& {C_\phi \over 2 \delta m_r} [(\del_m X^3)^2 + (\del_m \ph)^2]
\eeq
which is a free theory, a sigma model with the target space 
${\mathbb R}\times U(1)$.


\subsection{The fourth hierarchy: instanton inside the monopole inside the  vortex inside the wall}

Finally, we turn on the smallest parameters $\beta_r$ 
 in the hierarchy in Eq.~(\ref{eq:hierarchy}). 
The term proportional to the parameter $\beta_r$ perturbatively induces 
the following deformation 
term to the effective Lagrangian on the 
non-Abelian domain wall 
in Eq.~(\ref{eq:eff0}):
\beq
\Delta {\cal L}_{{\rm wall},\beta} 
 &=& \int dx^1 {\cal L}_{J,2}(H=H_{\rm wall}(x^1; X^1(x^i),U(x^i)))\non
 &=& \sum_r \frac{\pi \beta_r^2}{2 m''} \tr [X_r (U + U^\dagger)].
\eeq

Next, this term induces the following deformation term
to the effective Lagrangian of the non-Abelian vortex:
\beq
\Delta {\cal L}_{{\rm vortex},\beta} 
 &=& \int dx^2 \Delta {\cal L}_{{\rm wall},\beta}(U = U_{\rm vortex } (x^2;X^2(x^\alpha), \phi (x^{\alpha}) ))  \non
 &=& - \sum_r \frac{4\pi \beta_r^2}{m''{}^2} (\phi^\dagger X_r \phi).
\eeq
This is known as a momentum map (or a D-term in supersymmetric theories).
It  reduces in the $r$-th ${\mathbb C}P^1$ submanifold to
\beq
\Delta {\cal L}_{{\rm vortex},\beta} = 
 - \frac{4\pi \beta_r^2}{m''{}^2} 
 \frac{1}{1+|u|^2}
  \left(\begin{array}{cc}1 & u^*\end{array}\right)
 \sigma_1 
 \left(\begin{array}{c}1 \cr u\end{array}\right)
 = - \frac{4\pi \beta_r^2}{m''{}^2} \frac{u+u^*}{1+|u|^2}.
\eeq

Finally, this deformation induces the deformation term
 in the monopole effective action.
By considering a single monopole-string,
the effective Lagrangian 
is  given by 
\beq
\Delta {\cal L}_{r{\rm -th\; monopole},\beta} 
&=& \int dx^3 \Delta {\cal L}_{{\rm vortex},\beta} 
  (u = u_{\rm monopole} (x^3; X^3(x^m) ,\ph(x^m)) )  \non
&=& - C_r \cos \ph \equiv - V_{\rm monopole}  
  \non
   C_r &\equiv & \frac{ \pi^2 \beta_r^2}{2 \delta m_r m''{}^2}
 = \frac{\pi \beta_r^2 v^2}{4\delta m_r \gamma}.
\eeq
We thus arrive at the monopole effective Lagrangian summarized as 
\beq
 {\cal L}_{r{\rm -th \; monopole}, \beta} 
&=& {C_\phi \over 2 \delta m_r}
[(\del_m X^3)^2 + (\del_m \ph)^2  - D_r \cos \ph], \non
 D_r& \equiv & \frac{2 \delta m_r C_r}{C_\phi } 
= {\pi^{3/2}\over 2\sqrt{2}} {m \beta_r^2 \over v \sqrt{\gamma}}.
 \label{eq:SG}
\eeq
This is nothing but the sine-Gordon model.

We can consider multi-monopoles 
(multi-walls in the ${\mathbb C}P^{N-1}$ model). 
In that case, we will obtain 
a $U(1)^{N-1}$ coupled sine-Gordon model 
\cite{Fujimori:2016tmw} 
in which the interaction of sine-Gordon solitons 
on different domain walls was studied. 
 In our purpose, a single monopole is enough.

The monopole effective action in Eq.~(\ref{eq:SG}) admits 
sine-Gordon solitons. As the simplest,  
a single soliton solution 
is given by
\beq
 \ph = \ph_{\rm instanton}(x^4) 
 = 4 \arctan \exp{\sqrt{D_r \over 2} (x^4- X^4)} 
 + \pi . \ \quad
\eeq
The width of the soliton is $\Delta x^4 \sim 1 / \sqrt {D_r}
\sim \sqrt{v} \gamma^{1/4}/\sqrt{m} \beta_r$.
This carries a lump charge in 
the ${\mathbb C}P^1$ model as the vortex effective action 
\cite{Nitta:2012xq,*Kobayashi:2013ju} :
\beq
 T_{\rm lump} 
&\equiv& \int d^2x {i (\partial_i u^* \partial_j u - \partial_j u^* \partial_i u )
\over (1+|u|^2)^2} 
= 2 \pi k 
\eeq
with $k=1 \in \pi_2 ({\mathbb C}P^1)$,
which is also the topological charge 
$\pi_2 ({\mathbb C}P^{N-1})$ 
of the whole ${\mathbb C}P^{N-1}$.  
The lump charge in the vortex induces the instanton charge 
in the bulk, and thus
lumps in the vortex effective action correspond to 
Yang-Mills instantons in the bulk \cite{Hanany:2004ea,Eto:2004rz}.

Combining all discussions together, we finally reach the 
configuration in Fig.~\ref{fig:wvmi}.

\section{Relations among all topological solitons}\label{sec:relations}

\subsection{Subconfigurations}
Once we obtain the composite state of
the wall-vortex-monopole-instanton,
we can obtain various configurations 
consisting of a fewer number of solitons 
by removing some solitons from it.

\begin{figure}
\begin{center}
\includegraphics[width=1\linewidth,keepaspectratio]{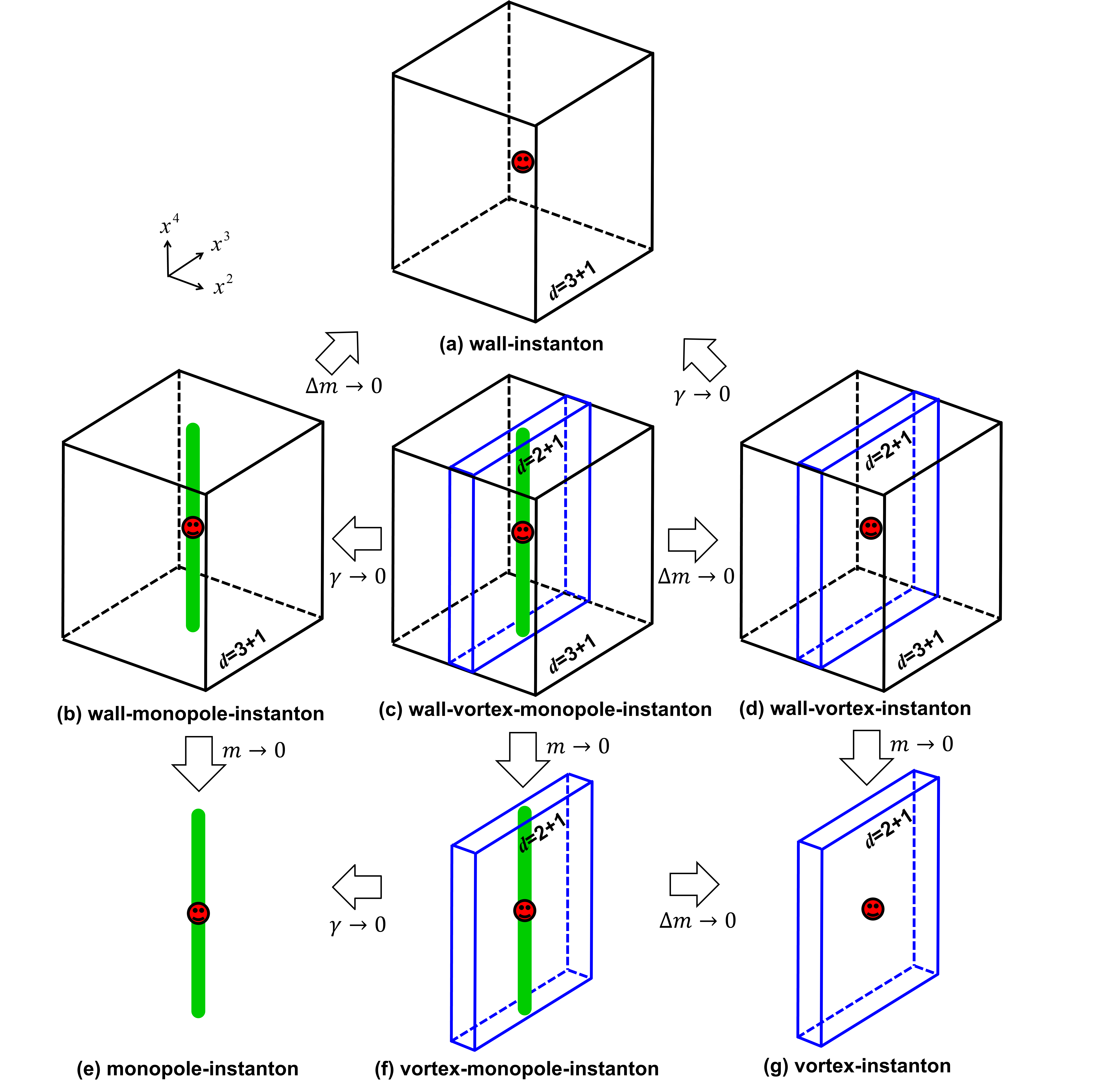}
\caption{Removing constituent solitons. 
Starting from (c) the wall-vortex-monopole-instanton, 
we can remove 
 the domain wall, vortex, or monopole in the limits 
$m\to 0$, $\gamma \to 0$ or $\delta m \to 0$, 
respectively 
to reach (f) the vortex-monopole-instanton, 
(b) wall-monopole-instanton or 
(d) wall-vortex-instanton, respectively.
We can remove two solitons
in the limits 
$\delta m, \gamma \to 0$, 
$m, \delta m \to 0$ or
$m, \gamma \to 0$  
to reach (a) the wall-instanton, 
(g) vortex-instanton or 
(e) monopole-instanton, respectively. 
We also can remove instantons from each configuration 
in the limit $\beta_a \to 0$.
\label{fig:all-limits}}
\end{center}
\end{figure}

Taking limits to send various parameters to zero, 
we obtain subconfigurations as schematically shown in 
Fig.~\ref{fig:all-limits}. 
We start from  (c) wall-vortex-monopole-instanton.  
We first remove one soliton from (c).
By taking the limit
$m\to 0$, $\gamma \to 0$, or $\delta m \to 0$, 
we can remove the domain wall, vortex, or monopole 
to reach (f) the vortex-monopole-instanton, 
(b) wall-monopole-instanton, and (d) wall-vortex-instanton, respectively.
Next, we remove two solitons from (c).
By taking the limit 
$\delta m, \gamma \to 0$, 
$m, \delta m \to 0$, or 
$m, \gamma \to 0$, 
we 
to reach (a) the wall-instanton, 
(g) vortex-instanton, or 
(e) monopole-instanton, respectively.
Only (e) is unstable as it is, but it can be stable 
if we make a closed loop \cite{Nitta:2013vaa}.
We also can remove instantons from each configuration 
in the limit $\beta_a \to 0$.

\subsection{Hidden relations}

All possible relations among topological solitons 
(domain walls, vortices, monopoles and Yang-Mills instantons) 
obtained thus far 
are summarized in 
Table \ref{table:relations}. 
Mother solitons can host solitons with less world-volume dimensions 
or equivalently of more codimensions 
as daughters.
The first row of the table shows that 
a $U(N)$ domain wall can host a $U(N)$ non-Abelian vortex, 
$SU(N)/U(1)^{N-1}$ monopole and $SU(N)$ instanton,
and in the domain wall effective theory, 
which is the $U(N)$ chiral Lagrangian, 
these solitons are realized as 
a $U(N)$ non-Abelian sine-Gordon soliton, 
$U(1)^{N-1}$ global vortex and 
$SU(N)$ Skyrmion, 
respectively.
In the second row, a vortex can host a monopole and instanton as 
a ${\mathbb C}P^{N-1}$ kink and ${\mathbb C}P^{N-1}$ lump,
respectively in the vortex world-volume theory, 
that is the ${\mathbb C}P^{N-1}$ model. 
Finally in the third row, a magnetic monopole-string 
can host an $SU(N)$ Yang-Mills instanton 
as a sine-Gordon soliton in the monopole world-volume 
effective theory which is the sine-Gordon model.

Table \ref{table:relations} contains further hidden relations 
among topological solitons. 

\begin{figure}
\begin{center}
\begin{tabular}{cc}
\includegraphics[width=0.25\linewidth,keepaspectratio]{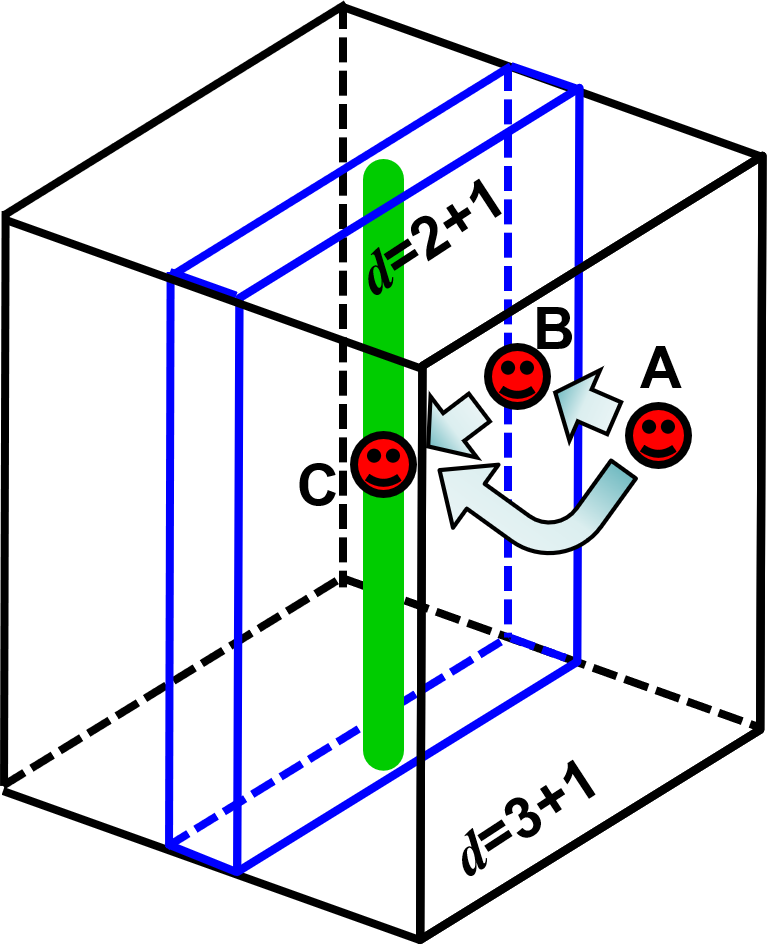}
& 
\includegraphics[width=0.25\linewidth,keepaspectratio]{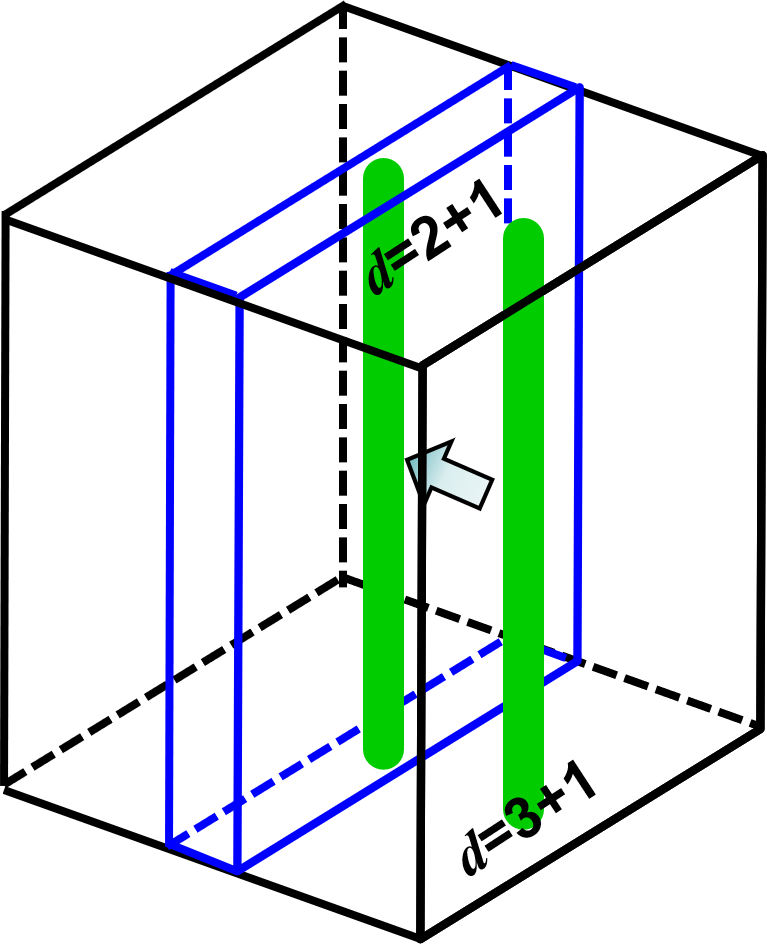}\\
(a) & (b)
\end{tabular}
\caption{Hidden relations.
(a) Within a domain wall, an $SU(N)$ Skyrmion at A turns to
 a ${\mathbb C}P^{N-1}$ lump at B 
 inside a $U(N)$ non-Abelian sine-Gordon soliton  (that is a non-Abelian vortex in the bulk point of view),
 and to a sine-Gordon soliton at C in a $U(1)^{N-1}$ global vortex 
 (that is a monopole in the bulk point of view). 
  Within a non-Abelian vortex, a ${\mathbb C}P^{N-1}$ lump at B becomes 
  a sine-Gordon soliton at C inside a ${\mathbb C}P^{N-1}$ kink.  
(b) A $U(1)^{N-1}$ global vortex (that is a monopole in the bulk point of view) becomes a  ${\mathbb C}P^{N-1}$ kink 
  inside a $U(N)$ non-Abelian sine-Gordon soliton.
\label{fig:hidden-relations}
}
\end{center}
\end{figure}

If one places an instanton inside a domain wall 
as A in Fig.~\ref{fig:hidden-relations}(a), 
it is a Skyrmion in the domain wall effective theory as denoted above. 
Then, if this Skyrmion moves into B on a non-Abelian vortex, 
it becomes a 
 ${\mathbb C}P^{N-1}$ lump in the vortex effective theory.
 This relation implies that 
   a $U(N)$ non-Abelian sine-Gordon in the $U(N)$ chiral Lagrangian 
   can host 
 an $SU(N)$ Skyrmion as a ${\mathbb C}P^{N-1}$ lump 
 in its world-volume effective theory which is the ${\mathbb C}P^{N-1}$ model \cite{Eto:2015uqa}.
 This corresponds to the first row and second column in  
Table \ref{table:relations2}.

If the Skyrmion at A moves to C on a monopole 
Fig.~\ref{fig:hidden-relations}(a) directly 
without passing through a vortex, 
it becomes a sine-Gordon soliton.
This implies that 
a $U(1)^{N-1}$ global vortex in the $U(N)$ chiral Lagrangian 
can host an $SU(N)$ Skyrmion as 
a sine-Gordon soliton in its world-volume effective theory 
which is a sine-Gordon model. 
 This corresponds to the second row and second column in  
Table \ref{table:relations2}. 
The case of $N=2$ was constructed in Ref.~\cite{Gudnason:2016yix}.
This relation for general $N$ is not studied yet.

We also can consider a monopole as a daughter soliton 
instead of the instanton.
As in Fig.~\ref{fig:hidden-relations}(b), 
if a monopole inside a domain wall is trapped into 
a non-Abelian vortex inside the domain wall as indicated 
by the arrow in the figure, 
it becomes a  ${\mathbb C}P^{N-1}$ kink.
This implies that 
  a $U(N)$ non-Abelian sine-Gordon in the $U(N)$ chiral Lagrangian 
   can host a $U(1)^{N-1}$ global vortex (which is a monopole in the bulk point of view)
 as a ${\mathbb C}P^{N-1}$ kink 
 in its world-volume effective theory which is the ${\mathbb C}P^{N-1}$ model. 
  This corresponds to the first row and first column in  
Table \ref{table:relations2}.
 This relation is not studied yet.

Finally, going back to Fig.~\ref{fig:hidden-relations}(a), 
if the instanton in the bulk (that is a ${\mathbb C}P^{N-1}$ lump (or baby Skyrmion)  inside the non-Abelian vortex) 
at B moves to C on a monopole in the bulk (that is a ${\mathbb C}P^{N-1}$ kink in the non-Abelian vortex), 
it becomes a sine-Gordon soliton, 
see Table \ref{table:relations3}.
This relation implies that 
a ${\mathbb C}P^{N-1}$ domain wall in 
the massive ${\mathbb C}P^{N-1}$ model can host a ${\mathbb C}P^{N-1}$ lump as a sine-Gordon soliton 
\cite{Nitta:2012xq,*Kobayashi:2013ju,Fujimori:2016tmw}. 
In the literature, this relation is known as a domain wall (baby) Skyrmion 
in field theory, attracting a lot of attention recently 
in condensed matter physics because of 
experimental confirmations
in chiral magnets \cite{PhysRevB.99.184412,*PhysRevB.102.094402,*Nagase:2020imn,*Yang:2021} 
(see also \cite{Kim:2017lsi}).

\section{Summary and Discussion \label{sec:summary} }

By using the effective field theory technique or the moduli approximation,   
we have constructed the wall-vortex-monopole-instanton 
configuration in Fig.~\ref{fig:wvmi}  
as the most general composite soliton
in the form of solitons within a soliton. 
It reduces to all the previously known composite solitons, 
providing relations among various 
topological solitons in various dimensions 
summarized in 
Tables \ref{table:relations}, 
\ref{table:relations2} and \ref{table:relations3}.
We have conjectured that this exhausts all possible 
relations among topological solitons.

We have considered non-degenerate masses 
$m_a \neq m_b$ if $a \neq b$, giving rise to 
Abelian domain walls in the ${\mathbb C}P^{N-1}$ model 
on a non-Abelian vortex as well as 
`t Hooft-Polyakov monopoles.
If we consider partially degenerate masses, 
the ${\mathbb C}P^{N-1}$ domain walls 
\cite{Eto:2008dm}
and 
monopoles on the vortex 
\cite{Nitta:2010nd}
are non-Abelian. 
In this case, 
instantons as 
sine-Gordon solitons on the monopole-string 
may be non-Abelian sine-Gordon solitons.

Among all major topological solitons in
 Table \ref{table:topological-solitons},  
we have not considered Hopfions. 
Hopfions inside a non-Abelian vortex was studied before
\cite{Nitta:2012mg}, but it is unclear to what it corresponds 
in the bulk.

As one of the most important applications, 
one can consider quantum effects, or 
non-perturbative effects in different dimensions.
Instanton corrections to monopoles correspond to 
vortex corrections to kinks through 
instanton vortex configurations. 
This gives a correspondence between 
a four dimensional $U(N)$ gauge theory 
and a two dimensional ${\mathbb C}P^{N-1}$ model. 
Hopefully, this relation can be generalized to 
other mother-daughter relations completed in this paper.

Electrically charged solitons or 
dyonic extensions can be also studied. 
As for dyonic extensions of individual solitons,  
dyonic instantons \cite{Lambert:1999ua}, 
dyons \cite{Julia:1975ff}, 
dyonic vortices \cite{Collie:2008za,*Eto:2014gya},
Q-lumps \cite{Leese:1991hr,*Abraham:1991ki} 
and 
Q-kinks \cite{Abraham:1992vb,*Abraham:1992qv} 
are known. 
Dyonic extensions of composite solitons are not 
well studied except for few examples:
domain wall networks and strings stretched 
between walls \cite{Eto:2007uc,Eto:2005sw}.

We have not studied fermions in this paper. 
When fermions are coupled to the system, 
topological solitons are often accompanied by 
fermion zero modes, such as on a kink
\cite{Jackiw:1975fn} and a vortex
\cite{Jackiw:1981ee}.
The existence of such fermions is ensured by 
the index theorem. 
For instance, 
the index theorem and fermion zero modes 
were studied for non-Abelian vortices in 
supersymmetric $U(N)$ gauge theory \cite{Hanany:2003hp}
and in dense QCD \cite{Yasui:2010yw,*Fujiwara:2011za}. 
However, they have not been well understood for 
composite solitons such as solitons within a soliton, 
and thus remain as one of future problems.

Probably, a more elegant framework to systematically understand 
composite solitons is offered by  
higher-form symmetry and higher group 
\cite{Gaiotto:2014kfa}. 
In fact, 
mathematical structures 
of the axion electrodynamics 
admitting composite solitons were recently clarified 
in terms of higher-form symmetries and higher groups  
\cite{Hidaka:2020iaz,*Hidaka:2020izy,*Hidaka:2021mml,*Hidaka:2021kkf}
(see also Refs.~\cite{Hidaka:2019jtv,*Hidaka:2019mfm}).
Mathematical structures behind the configuration in the present study 
are yet to be clarified. 

Finally, we hope that our most general configuration itself has a direct application in condensed matter physics. 
See Refs.~\cite{Volovik:2019goo,*Volovik:2020zqc} for composite solitons in helium-3 superfluids.

\section*{Acknowledgements}
The author thanks Minoru Eto and Sven Bjarke Gudnason for collaborations 
in various works. 
This work is supported in part by the JSPS Grant-in-Aid for Scientific Research (KAKENHI Grant No.~JP18H01217).

\begin{appendix}

\section{Mother and daughter of the same dimensions}\label{sec:same-dim}

As composite solitons in the form of 
solitons within a soliton, mother and daughter solitons can have the same dimensions in the following cases. 
In these cases, a single daughter soliton must be split into a set of 
fractional solitons. 

\begin{enumerate}
\item

{\bf Domain-wall sine-Gordon solitons}:

The double sine-Gordon model admits a false (metastable) vacuum in addition to the true vacuum in a certain parameter region 
\cite{PhysRevB.27.474,*Gani:2017yla} 
(see also Refs.~\cite{Eto:2018hhg,Eto:2018tnk,Ross:2020orc}).
 In this case, a single soliton profile has double peaks 
 which can be identified with a molecule of two constituents.
 Each constituent is an (unstable) domain wall or anti-domain wall separating 
 the true and false vacua, carrying half sine-Gordon soliton winding. 
 They are linearly confined by the false vacuum energy present between them. 
In the triple sine-Gordon model, a single sine-Gordon soliton 
is split into three constituents 
with 1/3 sine-Gordon topological charges 
\cite{Eto:2013hoa}.

\item

{\bf Vortex baby-Skyrmions (or vortex lump)}: 

In the baby Skyrme model ($O(3)$ or ${\mathbb C}P^1$ model) with the easy plane potential,  
a single baby Skyrmion is split into a pair of global vortex 
and anti-global vortex 
 with half baby-Skyrmion topological charges, 
constituting a half baby-Skyrmion molecule 
\cite{Jaykka:2010bq,*Kobayashi:2013aja,*Kobayashi:2013wra,*Leask:2021hzm}. 
Each half baby-Skyrmion (lump) 
can be interpreted as an Ising spin inside a global (anti-)vortex. 
Half baby-Skyrmions (lumps) are sometimes called merons.
For a ${\mathbb C}P^{N-1}$ model with 
a generalization of the easy plane potential, 
a single baby Skyrmion is split into a set of 
$N$ $U(1)^{N-1}$ global-vortex Skyrmions with $1/N$ lump ($\pi_2$)  topological charges \cite{Akagi:2021lva}.

A $U(1)$ gauged ${\mathbb C}P^1$ model (without a Skyrme term) 
or $U(1)^{N-1}$ gauged ${\mathbb C}P^{N-1}$ model
are  also known, 
for which vortices are local, and configurations are BPS,  
and thus can be embedded into a supersymmetric theory
\cite{Schroers:1995he,*Schroers:1996zy,*Baptista:2004rk,*Nitta:2011um}.
This can be applied to the stabilization of semilocal strings 
\cite{Eto:2016mqc}. 
See Ref.~\cite{Samoilenka:2017fwj} for the case with a Skyrme term.

\item
{\bf Monopole Skyrmions}: 

In a Skyrme model with the potential term admitting 
$S^2$ vacua, 
a single $SU(2)$ Skyrmion is split into a pair of a global monopole
and an anti-global monopole
 with half-Skyrmion topological charges, 
constituting a half Skyrmion molecule 
\cite{Gudnason:2015nxa}.   
Each half Skyrmion can be interpreted as an Ising spin inside a global (anti-)monopole.

An $SU(2)$ gauged version is also known, 
for which monopoles are local (`t Hooft-Polyakov type)
\cite{Arthur:1996np,*Kleihaus:1999ea,*Grigoriev:2002qc}.
It is also called a Skyrmed monopole.

\end{enumerate}

\section{Solitons on compactified spaces with twisted boundary conditions}\label{sec:tbc}

Topological solitons 
on compactified spaces
${\mathbb R}^{d-1} \times S^1$ 
with twisted boundary conditions along $S^1$
have half topological charges 
together with topological charges in ${\mathbb R}^{d-1}$. 
These solitons are also composite solitons 
in the form of solitons within a soliton.

\begin{enumerate}

\item
{\bf Monopole instantons}:  

Yang-Mills instantons on ${\mathbb R}^3 \times S^1$ 
are called calorons.
The situation with twisted boundary condition 
is known as 
calorons with a nontrivial holonomy along 
$S^1$~\cite{Lee:1998vu,*Lee:1998bb,*Kraan:1998pm,*Kraan:1998sn}.
A single $SU(2)$ instanton can be decomposed into 
a pair of a monopole string 
and an anti-monopole string winding along $S^1$ 
with half instanton charges. 
A bion is a pair of a monopole-string with 
$+1/2$ instanton charge 
and an anti-monopole string with $-1/2$ instanton
charge, 
and has been considered for applications 
to confinement and mass gap 
\cite{Unsal:2007jx,*Anber:2011de,*Argyres:2012ka}.
Similarly, an $SU(N)$ instanton is decomposed into 
$N$ monopole strings with $1/N$ instanton charges.

\item
{\bf Vortex Skyrmions}: 

Skyrmions on ${\mathbb R}^2 \times S^1$ 
with twisted boundary conditions 
are considered with 
\cite{Harland:2008eu} 
and without
\cite{Nitta:2015tua} 
the Skyrme term.
A single $SU(2)$ Skyrmion can be decomposed into 
a pair of a global vortex string 
and an anti-global vortex string winding along $S^1$ 
with half Skyreme ($\pi_3$) charges. 
A bion is a pair of a global vortex-string with 
$+1/2$ Skyrme charge 
and an anti-global vortex string with $-1/2$ Skyrme
charge \cite{Nitta:2015tua}. 
Likewise, a single $SU(N)$ Skyrmion is decomposed into 
$N$ $U(1)^{N-1}$ global-vortex Skyrmions 
with $1/N$ Skyrme charges 
\cite{Nitta:2015tua}.  
Applications of bions in this case are yet to be clarified. 
This relation can be embedded inside 
a non-Abelian domain wall, 
reproducing the aforementioned monopole instantons 
\cite{Nitta:2015mxa}.

\item
{\bf Domain-wall lumps}:  

\begin{enumerate}

\item
${\mathbb C}P^{N-1}$ lumps  on ${\mathbb R}^1 \times S^1$ 
with twisted boundary conditions were first found in Ref.~\cite{Eto:2004rz}  
(see also Refs.~\cite{Bruckmann:2007zh,*Brendel:2009mp,*Harland:2009mf,*Bruckmann:2014sla}).
See Ref.~\cite{Harland:2007pb} for baby Skyrmions 
for the case with the Skyrme term.
A single ${\mathbb C}P^1$ lump can be decomposed into 
a pair of a domain wall  
and an anti-domain wall winding along $S^1$ 
with half lump ($\pi_2$) charges.
Similarly a single ${\mathbb C}P^{N-1}$ lump can be decomposed into 
a set of $N$ domain walls  
with $1/N$ lump ($\pi_2$) charges.
A bion is a pair of a domain wall with 
$+1/2$ lump charge 
and an anti-domain wall with $-1/2$ lump charge, 
and similar for the ${\mathbb C}P^{N-1}$ model \cite{
Dunne:2012ae,*Dunne:2012zk,*Misumi:2014jua,*Misumi:2014rsa,*Fujimori:2018kqp},
which has been extensively applied to resurgence theory 
of the ${\mathbb C}P^{N-1}$ model.
This relation can be embedded inside 
a non-Abelian vortex
reproducing the aforementioned monopole instantons 
\cite{Eto:2004rz}. 
Recently,  
Monte Carlo simulations have been performed  \cite{Misumi:2019upg,*Fujimori:2020zka} 
in which fractional lumps and bions have been observed.

\item
Grassmann lumps on ${\mathbb R}^1 \times S^1$ 
with twisted boundary conditions were first found in 
Refs.~\cite{Eto:2006mz,*Eto:2007aw}.
Bions and application to the resurgence theory can be found in Refs.~\cite{Misumi:2014bsa,*Dunne:2015ywa}.

\end{enumerate}

\item
{\bf Lump-string Hopfions}: 
Hopfions on ${\mathbb R}^2 \times S^1$ were 
studied in Ref.~\cite{Kobayashi:2013aza}.
\end{enumerate}
\end{appendix}

\bibliographystyle{apsrev4-1}

\end{document}